\documentclass[final,5p,times,twocolumn,nopreprintline]{elsarticle}
\biboptions{sort&compress}

\usepackage{amssymb}
\usepackage{amsmath}
\usepackage{graphicx}
\graphicspath{{./}{Figures/}}
\usepackage{dblfloatfix}
\usepackage{gensymb}
\usepackage{cleveref}
\usepackage{xcolor}
\usepackage{soul}
\usepackage[normalem]{ulem}
\usepackage{float}
\usepackage{url}

\providecommand{\href}[2]{#2}
\DeclareUrlCommand\path{\urlstyle{same}}

\begin{document}

\makeatletter
\let\PrelimSaved@textbottom\@textbottom
\makeatother
\raggedbottom

\begin{frontmatter}

\title{Shock Wave Propagation in a Conduit with a Porous Wall}

\author[inst1]{Omri Ben Haim}

\affiliation[inst1]{organization={Technion - Israel Institute of Technology, Faculty of Mechanical Engineering},
    city={Haifa},
    postcode={3200003},
    country={Israel}}

\author[inst1]{Yoav Gichon}
\author[inst1]{Raz Heppner}
\author[inst1]{Omri Ram\corref{cor1}}\ead{omri.ram@technion.ac.il}
\cortext[cor1]{Corresponding author}

\begin{abstract}
A shock traveling along a porous wall can generate a transmitted front that overtakes it. We investigate how wall geometry and cavity height govern the relative motion of these fronts and the resulting pressure loading when the porous wall separates a conduit from a cavity. We use time-resolved schlieren imaging and pressure measurements to test six wall geometries at cavity heights of $9$, $39$, and $69$~mm and an incident Mach number of $1.45$. Complementary head-on experiments and two-dimensional simulations examine how the wall passages shape shock transmission. Increasing the cavity height from $9$ to $69$~mm increases the cavity front's arrival lag behind the conduit shock near the entrance from approximately $60$ to $185$~$\mu\mathrm{s}$ and shifts the peak cavity loading at front arrival downstream. In taller cavities, the transmitted front travels farther before reflecting from the outer wall, and the resulting Mach stem spans the cavity farther downstream. At the two larger heights, wall geometry determines whether the cavity front overtakes the conduit shock. The cavity front becomes nearly normal closest to the entrance for circular elements and farthest downstream for square elements and aligned triangles with their apex facing the cavity. These latter two geometries also sustain stronger conduit shocks. In head-on tests at fixed void fraction and frontal blockage, reversing the aligned triangles so that their apex faced the cavity narrowed the passage entrances and increased the characteristic pressure-rise time from approximately $175$ to $290$~$\mu\mathrm{s}$. The slower pressure buildup is consistent with the cavity front becoming nearly normal farther downstream for this geometry when the shock travels along the wall.

 \end{abstract}





\end{frontmatter}

\makeatletter
\let\@textbottom\PrelimSaved@textbottom
\makeatother

\section{Introduction}
\label{sec:Intro}
Shock waves form in many engineering systems and industrial processes. Examples include sudden releases of high-pressure gas from piping and exhaust valves \cite{duan2015experimental, winroth2019shock}, explosive combustion \cite{shepherd2009structural}, and high-speed propulsion systems \cite{arad2024simulation}. In all of these, the shock waves propagate in confined passages. Changes in passage geometry, such as contractions and expansions, alter the wave fronts and downstream pressure histories through reflection, diffraction, and interaction with the induced flow \cite{gichon_dynamics_2024,heppner2026shock}. The resulting pressure loads depend on the incident wave and the passage geometry \cite{kim2004study} and can threaten the structural integrity of the confinement \cite{shepherd2009structural}. The hazard is not limited to small systems: larger confined structures, such as buildings, ventilation networks, and mine shafts, are also exposed to shock-induced loads \cite{ram2016dependence,kokkinakis_internal_2023,wang2021experimental}. Furthermore, shock waves that emerge from their confinement expose the surroundings to blast loading and impulsive noise \cite{gan2024blast,sasoh1998attenuation}. These hazards motivate the development and assessment of measures to attenuate shock waves and reduce their associated loads.

Porous barriers have demonstrated the potential to mitigate loads that develop behind them following the impingement of shock waves \cite{andreopoulos2007moving, berger2010experimental, Chaudhuri2013, hadjadj_shock_2013, frolov_effectiveness_1993, vinod2025attenuation}. Previous research has examined a variety of porous barriers, including rigid ceramic foams \cite{Levy1993,ram2013simple,ram2015analysis}, arrays of solid obstacles \cite{Chaudhuri2013,sha_influences_2014}, barriers and grids \cite{berger2010experimental,britan2006gridsorifice,berger2015experimental,berger2015numerical}, granular filters \cite{britan2001granular}, and perforated plates \cite{britan2004perforated,skews2005shock,vinod2024predicting,vinod2025attenuation}. These studies use experiments, numerical simulations, or both to examine transmitted shocks and downstream loading.

In cases of head-on impingement of a shock wave on a porous barrier, the geometry and arrangement of the barrier elements influence both the passage of the shock through the wall and the ensuing flow. When the shock reaches the barrier, portions of the incident front reflect from the upstream-facing surfaces of the elements, while others propagate through the passages between neighboring elements. The reflected portions subsequently combine into an upstream-propagating shock, whereas the transmitted front advances through passages whose cross-sectional area varies with the element shape. Reflections within these passages generate transverse waves, and the transmitted shock diffracts as it encounters changes in passage geometry. Meanwhile, the shock-induced flow develops boundary layers and can separate off sharp edges or farther along the element surfaces, generating shear layers and vortices that interact with the wave system \cite{Chaudhuri2013,heppner2026shock}.

The location and development of these interactions differ between element shapes. In the study of single-column blockage array configurations by Chaudhuri et al. \cite{Chaudhuri2013}, upstream-pointing triangles and circular elements produced a gradual narrowing of the passages, whereas squares and downstream-pointing triangles produced an abrupt constriction. For upstream-pointing triangles, the initial regular reflection evolved into a Mach reflection as the shock progressed along the inclined surfaces, and the reflected wave was the weakest among the geometries examined. For squares and downstream-pointing triangles, vortices developed near the abrupt passage entrances and interacted with the internal transverse waves. At a later stage, the interaction of transverse waves with the transmitted front differed particularly for downstream-pointing triangles and was associated with a lower transmitted-front speed. These observations show how the geometry and orientation of the blockage elements change the reflection pattern and the subsequent coupling between the waves and the induced flow.

Further studies examined how specific geometric features modify these processes. Sha et al. \cite{sha_influences_2014} showed that the upstream-facing slope, upper-side width, and downstream-facing slope have distinct influences on shock attenuation, with the downstream-facing slope having a weaker influence than the other two features in the configurations studied. For inclined barriers, Berger et al. \cite{berger2015experimental} related changes in reflected-wave strength and downstream flow to the converging or diverging character of the opening and the development of vortices near its edges. Their investigation of double barriers further showed that the influence of the second barrier depended on its separation from the first \cite{berger2015numerical}. Thus, both the local passage geometry and the spacing between successive obstacles affect the attenuation process. In their comparisons of obstacle matrices, Chaudhuri et al. \cite{Chaudhuri2013} found that a staggered arrangement of downstream-pointing triangular elements, with their bases facing the incident shock, provided the most effective attenuation among the configurations examined, considering both static and dynamic pressure impulses. This result motivates examining the influence of element orientation and arrangement when the shock travels along the barrier instead of impinging on it, as in the present study.

The downstream boundary condition further modifies the pressure response behind the barrier. Before reflections from downstream boundaries return, the transmitted shock propagates along the conduit, while the pressure difference across the barrier drives flow through its openings. Studies of this stage examine the attenuation of the transmitted shock and the influence of barrier geometry on the downstream wave and flow field \cite{andreopoulos2007moving,berger2015experimental,berger2015numerical,britan2004perforated}. When an end wall is positioned at a finite stand-off distance behind the barrier, the transmitted shock reflects from it and returns toward the barrier, producing successive wave reflections within the intervening gas volume \cite{ram2013simple,ram2015analysis,ram2018pressure}. Meanwhile, pressure-driven mass flow through the barrier changes the pressure within this confined volume, which in turn modifies the pressure difference driving the flow. Under sustained upstream loading, this coupled process tends to reduce the pressure difference between the upstream region and the confined volume. Predicting the resulting pressure build-up therefore requires a relation between the pressure difference and the mass flow rate through the barrier, which can be nonlinear when compressibility and inertial effects are significant.

To describe the coupled evolution of flow through the barrier and pressure within the confined volume, several studies have developed macroscale models relating the pressure imposed on the upstream face of a porous barrier to the pressure developing at an end wall behind it \cite{ram2013simple,ram2015analysis,ram2018pressure,vinod2024predicting}. These models represent the porous barrier and the gas confined within the barrier and in the volume behind it as a lumped system, without resolving the detailed flow around individual elements. The resistance to flow through the barrier, together with the compressibility of the confined gas, determines how the downstream pressure responds to the imposed loading. This response can exhibit low-pass filtering behavior, reducing rapid pressure variations and producing a smoother pressure rise at the end wall. For an imposed pressure step, the pressure build-up can be described using a characteristic response time that depends on the barrier properties, the confined gas volume, and the gas properties \cite{ram2013simple}. The approach has also been extended to time-dependent loading and examined in the frequency domain \cite{ram2015analysis,vinod2024predicting}. Related models have been applied to rigid porous SiC filters \cite{ram2013simple,ram2015analysis} and arrays of perforated plates \cite{ram2018pressure,vinod2024predicting}. These applications motivate examining whether the same modeling framework can describe other porous configurations, provided that their flow resistance and confined gas volume are appropriately represented.

In contrast to head-on impingement, a shock wave propagating along a porous wall loads successive portions of its surface as the front advances downstream. The resulting flow through the wall modifies the shock in the main passage and generates disturbances on the opposite side. Frolov and Gelfand \cite{frolov1992confined} described the attenuation of shocks traveling through perforated channels using a quasi-one-dimensional model that accounts for lateral outflow through the openings. Experimental and numerical investigations by Onodera and Takayama \cite{onodera1994analysis,onodera1998shape} related this outflow to a discharge coefficient, while subsequent work examined how it determines the deformation of the shock front over the perforated wall. The conditions behind the openings also influence the interaction. Sasoh et al. \cite{sasoh1998attenuation} studied weak shocks propagating along pseudo-perforated walls, in which individual openings were backed by closed cavities. They showed that diffraction and reflection within these cavities modify the pressure field in the main passage, with local pressure amplification occurring despite attenuation of the leading shock.

These studies establish the importance of flow through the wall and the disturbances generated behind it in determining the evolution of the incident shock. For a porous wall separating a conduit from a continuous adjacent cavity, the response involves fronts propagating along both volumes while successive portions of the wall are exposed to the incident pressure rise. The internal wall geometry determines the passages through which the transmitted disturbances and subsequent flow enter the cavity, while the cavity height sets the distance that these disturbances travel before they encounter the opposing boundary. For this configuration, the question is how the geometry of the wall passages and the height of the cavity affect the development of the cavity front, its position relative to the conduit shock, and the pressure rise produced by each front as it travels downstream.

To address this question, we examine six porous wall geometries at three cavity heights. The walls comprise circular cylinders, square elements, and aligned or staggered triangular elements with their apex facing either the conduit or the cavity. Schlieren recordings are used to track the fronts in both volumes and to follow the growth of the Mach stem across the cavity through the trajectory of the triple point. The relative arrival of the conduit and cavity fronts is evaluated from both the schlieren recordings and paired pressure measurements. The pressure records also show how the pressure peak at front arrival varies downstream and how much of the late-time cavity pressure is reached at front arrival near the cavity entrance.

Complementary head-on experiments examine how the same wall geometries affect pressure build-up behind the barrier and provide a characteristic response time for each wall. Two-dimensional numerical simulations of head-on impingement on the four aligned walls complement these measurements by resolving the wave pattern and the flow within the wall passages, which the experiments cannot access. We then test whether the head-on calibration describes the cavity pressure-rise timescale near the cavity entrance under parallel loading, with the late-time pressure prescribed from the measurements.

\section{Methodology}
\subsection{Problem Description}
Figure~\ref{fig:Experimental}\textit{(a,b)} illustrates the two investigated configurations. Both configurations have a main conduit height of $E=40$~mm, a porous-wall thickness of $L=21$~mm, and the same set of internal wall geometries. In the head-on configuration, shown in Figure~\ref{fig:Experimental}\textit{(a)}, the incident shock impinges on the porous wall, with a fixed end wall positioned $39$~mm behind its rear face. This configuration was also simulated numerically for the four aligned geometries (Section~\ref{sec:numerics}). In the parallel configuration, shown in Figure~\ref{fig:Experimental}\textit{(b)}, the shock propagates along the porous wall, which separates the main conduit from an adjacent cavity. The cavity height $H$ is measured from the rear face of the porous wall to the cavity outer wall (labeled ``Cavity wall'' in Figure~\ref{fig:Experimental}\textit{(b)}). Three cavity heights were investigated, with $H_1=9$~mm, $H_2=39$~mm, and $H_3=69$~mm. The corresponding step height from the conduit face of the porous wall to the cavity outer wall is $L+H$. The head-on stand-off distance equals the intermediate cavity height $H_2$, so that the two configurations can be compared at the same distance between the porous wall and the boundary behind it.

\subsection{Experimental Setup}
\begin{figure*}[t]
    \centering
    \includegraphics[width=\textwidth]{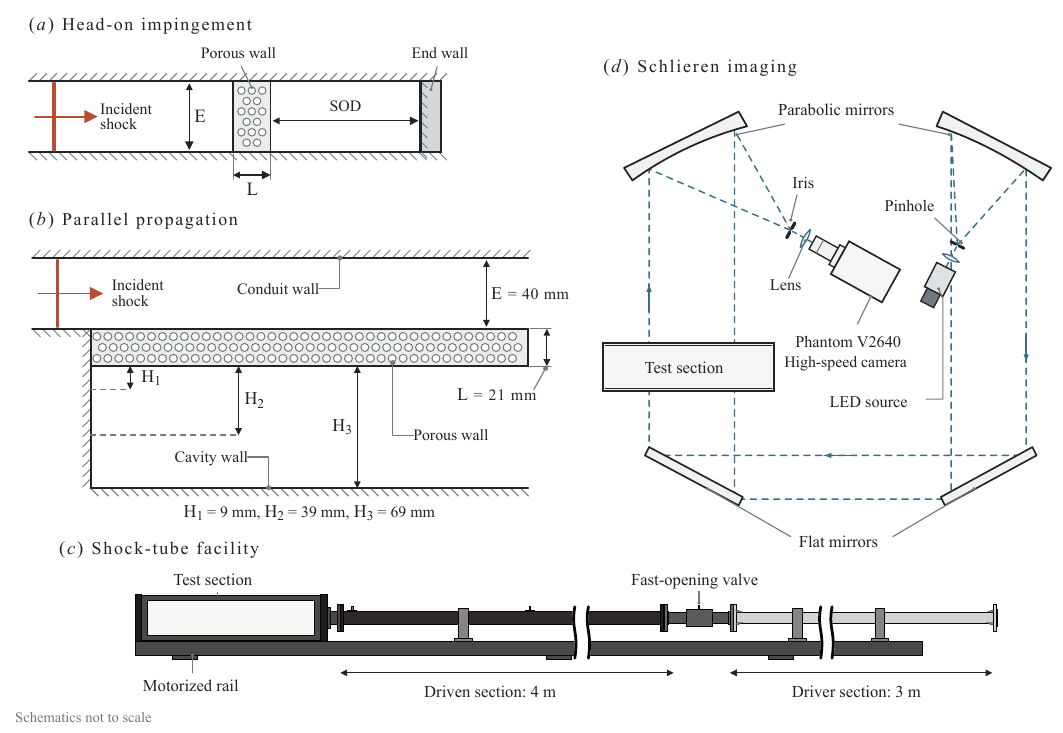}
    \caption{Schematics of the experimental system: (a) the head-on impingement test section; (b) the parallel-propagation test section; (c) the shock-tube apparatus; (d) the optical setup used for high-speed schlieren imaging.}
    \label{fig:Experimental}
\end{figure*}
The experiments were conducted in the Transient Fluid Mechanics Laboratory (TFML) shock tube facility at the Technion -- Israel Institute of Technology. Figure~\ref{fig:Experimental}\textit{(c)} depicts the shock-tube apparatus, which comprises a $3000$~mm long cylindrical driver section with an inner diameter of 52.5~mm, a 4080~mm long driven section with a $40 \times 40$~mm square cross-section, and an interchangeable rectangular test section. An ISTA KB-40 pneumatic fast-opening valve separates the driver and driven sections. The valve is designed to open within 1 to 2~ms, producing repeatable shock waves with incident Mach numbers ($M$) ranging from 1.2 to 1.8, with a variation of less than 0.3\% between consecutive experiments.

Two test sections were used, one for each configuration. The first test section, used for the head-on impingement experiments, was 1000~mm long with a $40\times40$~mm square cross-section and was fitted with an interchangeable porous model and a permanent end block. Both the porous model and the end block were held by mortises and grooves machined into the aluminum wall of the test section, which restricted their movement on impact. Two pressure transducers were flush-mounted on the conduit wall 105~mm and 6~mm upstream of the porous wall to monitor the incoming shock speed, and one transducer was mounted in the center of the end block. The second test section, used for the parallel-propagation experiments, was 1000~mm long and 40~mm wide, with an open downstream end. In this configuration, the porous wall separates two volumes: the main conduit and a cavity of variable height. This test section was equipped with 14 flush-mounted pressure transducers, which were divided equally between the conduit wall and the cavity outer wall and installed at matching streamwise positions. On each wall, the seven transducers were mounted at constant intervals of $105$~mm, with the first transducer located $54$~mm from the entrance of the test section, which was taken as $x=0$. These measurement positions are referred to below as stations~1 to~7.

The pressure histories were measured using ENDEVCO model 8510B-500 pressure transducers coupled with an ENDEVCO model 136 amplifier and recorded on four LeCroy Waverunner LT344 digital oscilloscopes. The transducers have a 3.8~mm sensing face and a resonance frequency of 500~kHz, and the amplifier response is flat within $\pm5\%$ up to 50~kHz, with a $-3$~dB point at 200~kHz. Each channel was sampled at 20~MHz in the parallel experiments, and at 10~MHz in the head-on experiments, with $10^5$ samples per channel, giving records of 5 and 10~ms, respectively. The measured 10--90\% rise time of the incident shock, about 8~$\mu$s, matches the time the shock takes to cross the transducer face, so the transducer size, rather than the sampling rate, limits the resolution of the pressure jumps. Two additional transducers, located $1000$~mm apart in the driven section upstream of the test section entrance, were used to determine the incident shock Mach number.

The imaging and data acquisition systems were synchronized via a Quantum Composers 9400 external timing box, triggered by the arrival of the shock wave at the first upstream transducer.

High-speed schlieren imaging was employed to visualize and analyze the flow features. The schematic in Figure~\ref{fig:Experimental}\textit{(d)} presents the schlieren setup. Light from a Thorlabs MCWHL5-C1 LED is directed through a pinhole and collimated by an Edmund Optics 32-276-533 parabolic mirror with an unobstructed diameter of $286$~mm. The collimated beam is then aligned to propagate perpendicularly through the test section using two flat mirrors. A second parabolic mirror focuses the light through an iris at its focal point. The iris opening is set to create a schlieren image sensitive enough to capture the density variations in the flow that result from the expansion of the shock waves into the cavity and their interaction with the porous wall. A Phantom V2640 camera, equipped with an imaging lens, recorded the flow field at up to 80~kHz with a resolution of $1024 \times 256$ pixels. To allow for imaging at various locations without realigning the optics, the entire shock tube was mounted on a stepper-motor-driven movable platform. The facility was fully automated via a custom LabVIEW controller. The image scale was calibrated at each viewing position using a calibration plate.

To facilitate optical visualization, both test sections were equipped with $30$~mm thick clear PMMA windows. These were bolted to the structure and fitted with seals to ensure a leak-proof assembly. For the second test section, the windows were grooved along the cavity, forming a mortise for mounting a porous wall sample covering the entire length of the cavity.

The driven and test sections were filled with air at laboratory conditions before each experiment. The parallel test section was open downstream, whereas the head-on configuration terminated at a fixed end wall. About 10 minutes were allowed between experiments for the conditions in the driven and test sections to equalize with the laboratory air, ensuring that the incident shock wave propagates into quiescent air with uniform atmospheric properties.
\subsection{Numerical Scheme}
\label{sec:numerics}
The two-dimensional compressible Navier--Stokes equations were solved using the in-house Large Eddy Simulation solver described and validated in \cite{heppner2026shock}. Air is treated as a calorically perfect gas with $\gamma = 1.4$, a gas constant of $287$~J\,kg$^{-1}$\,K$^{-1}$ and a Prandtl number of $0.71$. The primitive variables are transformed to characteristic variables and reconstructed at the cell faces using the fifth-order monotonicity-preserving scheme (MP5) of Suresh and Huynh~\cite{suresh1997accurate}, and the inviscid fluxes are evaluated using the HLLC approximate Riemann solver \cite{harten1983upstream,Toro1997}. The viscous fluxes are evaluated at the cell faces from sixth-order explicit central-difference gradients with $\alpha$-damping ($\alpha = 4$) \cite{chandravamsi2024high}. Time integration is performed using a three-stage TVD Runge--Kutta scheme \cite{shu1988efficient} with a CFL number of $0.5$.

Head-on impingement on the four aligned geometries (circular, square, and the two triangular orientations) was simulated. The computational domain was an 8~mm high channel bounded by the centerlines of two neighboring element columns, so that half elements protrude from both lateral boundaries and the 2~mm gap lies at its center. The domain extends 10~mm upstream of the front face of the wall and 39~mm downstream of its rear face, where a slip wall represents the end wall. No-slip adiabatic conditions are applied on the element surfaces and on the lateral boundaries, and the flow variables are extrapolated at the upstream boundary, which allows the reflected shock to leave the domain. The multi-block structured grids contain between $0.33\times10^{6}$ and $0.74\times10^{6}$ cells, with a streamwise spacing of approximately 28~$\mu$m and at least 70 cells across the minimum gap. Each simulation is initialized with a planar shock of $M = 1.4$ located 5~mm upstream of the wall and propagating into quiescent air at $101.3$~kPa and $293$~K.

\subsection{Porous Samples }
\begin{figure}[!tp]
    \centering
    \includegraphics[width=0.75\columnwidth]{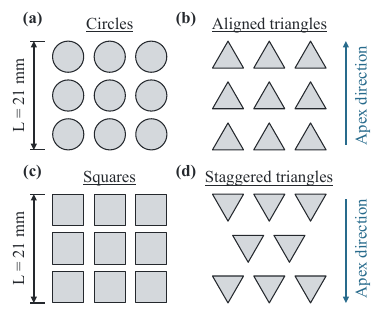}
    \caption[]{Illustration of the internal geometries and their arrangement inside the porous wall: (a) circles; (b) aligned triangles; (c) squares; (d) staggered triangles. The double arrows indicate that the triangular elements were tested with the apex facing either the conduit or the cavity.}
    \label{fig:PorousModels}
\end{figure}

To examine the effect of the internal structure of the porous wall, six porous models were manufactured from PLA on an FDM 3D printer (Figure~\ref{fig:PorousModels}). Each model consisted of three rows of identical elements with a circular, square, or triangular cross-section. The characteristic length of all shapes was 6~mm, namely the circle diameter, the square side, and the triangle base. Adjacent elements were separated by a clear gap of 2~mm along the wall's length, and the three rows were equally spaced across its thickness $L$. The void fraction $\phi$ (Table~\ref{tab:headon}) is the open fraction of the layer cross-section, $40 \times 21$~mm$^2$ in the head-on wall, which holds five elements per row, or four in the shifted row of the staggered walls.

The triangular models were tested in two orientations, once with their apex facing the main conduit (apex-to-conduit) and once with the apex facing the cavity (apex-to-cavity). In the head-on configuration, the same names denote the apex facing the incident shock and the end wall, respectively. To investigate the effect of the internal arrangement, another set of triangular models was manufactured with the second row shifted by half a pitch, so that its elements are centered on the gaps of the first row. This arrangement is referred to as staggered, and the unshifted one as aligned. In addition to the internal geometry that interacted with the flow, each model had tenons extending in two directions to fit into the mortises that were machined into the test section.

\subsection{Simplified Model for Pressure Prediction}
\label{sec:model}
Ram et al.~\cite{ram2013simple,ram2015analysis,ram2016dependence,ram2018pressure} presented a macroscale model for the pressure build-up behind a porous barrier after head-on impingement by a shock wave. The model considers the end-wall configuration and treats the porous barrier, the gas within it, and the confined volume behind it as a single input-output system. The input is the near-step pressure imposed on the front face of the porous layer, and the output is the pressure build-up measured on the end wall of the confined volume. Assuming approximately isentropic filling, Ram and Sadot \cite{ram2013simple} showed that the normalized pressure histories could be collapsed using a time scaling based on the effective length of the gas-filled region. For a fixed cross-sectional area, this length accounts for both the gas within the pores and the gas in the cavity behind the layer. Following the time-constant notation of Ram et al.~\cite{ram2018pressure}, the characteristic response time and corresponding pressure response are expressed as
\begin{equation}
    \tau = \frac{\left(\mathrm{SOD}+L_{\mathrm{eff}}\right)^{\gamma}}{\alpha},
    \qquad
    \frac{P(t)}{P_{imp}} = 1-e^{-t/\tau},
    \label{eq:Ram_Model}
\end{equation}
where $\tau$ is the characteristic pressure-rise time, $\mathrm{SOD}$ is the stand-off distance between the back face of the layer and the end wall, $L_{\mathrm{eff}} = \phi L$ is the effective barrier length obtained from the void fraction $\phi$ and the layer thickness $L$, and $\gamma$ is the ratio of specific heats. For plotting the head-on pressure histories, we retain the length-scaled time $t_s=t/(\mathrm{SOD}+L_{\mathrm{eff}})^\gamma$, with units of $\mu$s\,mm$^{-\gamma}$; thus $\alpha t_s=t/\tau$. Here, $P(t)$ denotes the end-wall overpressure relative to the initial ambient pressure, and $P_{imp}$ denotes the asymptotic overpressure imposed on the upstream face of the porous wall. Time is measured from the arrival of the transmitted front at the end-wall transducer. Within this model, a calibration experiment determines an effective coefficient $\alpha$ for a given sample, which lumps all macroscopic properties of the porous layer (porosity, tortuosity, pore shape and friction). In the present work, its applicability beyond the head-on calibration condition is tested by comparison with parallel-loading measurements.

\section{Results}
\label{sec:Results}

This section presents experimental results for the two configurations, complemented by numerical simulations of head-on impingement. Section~\ref{sec:headon} calibrates the lumped model of \cite{ram2013simple, ram2015analysis, ram2018pressure} for the six porous wall geometries under head-on impingement, and uses the simulations to examine the wave propagation within the wall. Section~\ref{sec:parallel} examines how wall geometry and cavity height affect shock-front evolution and pressure histories in the conduit and cavity under parallel propagation, and assesses whether the head-on calibration describes the cavity pressure-rise timescale in this configuration.

\subsection{Head-on Impingement}
\label{sec:headon}

In the head-on scenario, which has been described thoroughly in \cite{ram2013simple,ram2015analysis, ram2018pressure,skews2005shock,britan2001granular, kitagawa2006attenuation}, the unified transmitted shock front reflects normally off the end wall and then reverberates between the end wall and the rear face of the porous wall. Some of the reverberating shocks emerge through the wall into the main conduit and join the reflected shock wave. Through these repeated interactions, the transmitted shock weakens and subsides over time.

Section~\ref{sec:Head-onSimulation} uses the simulations to follow the shock through each wall, and Section~\ref{sec:Head-onCalibration} presents the measured end-wall pressures and the calibration of the model.

\subsubsection{Shock transmission and flow evolution within the porous wall}
\label{sec:Head-onSimulation}

The head-on experiments record the pressure upstream of the porous wall and on the end wall, but they do not resolve the wave pattern or the flow within the passages between the elements. To complement these measurements, head-on impingement on the four aligned geometries was simulated numerically.

\begin{figure*}[t]
    \centering

    \includegraphics[width=\textwidth]{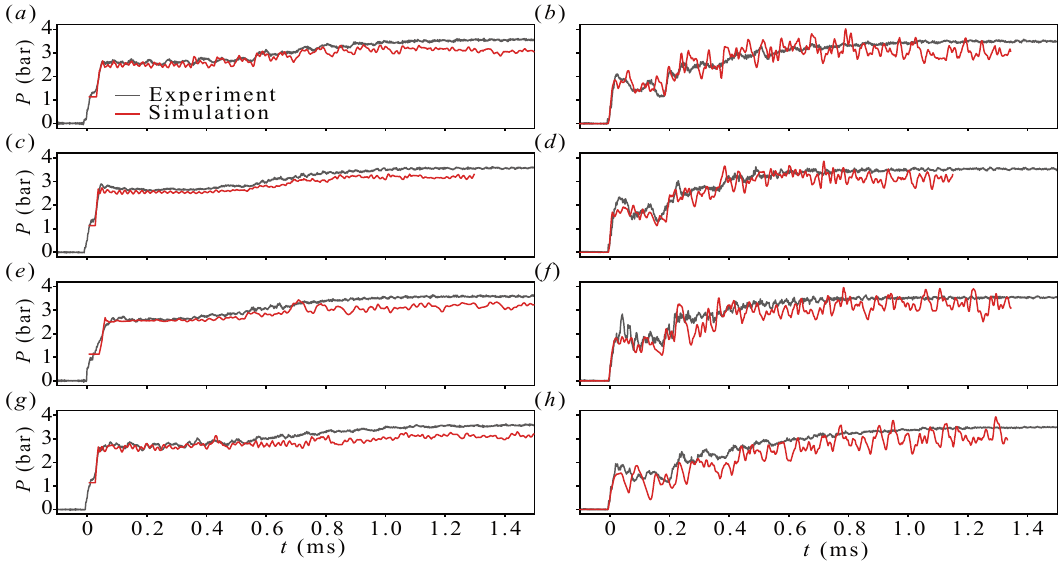}
    \caption{Measured and computed pressure histories for head-on impingement, (a,c,e,g) 6~mm upstream of the porous wall and (b,d,f,h) at the end wall: (a,b) circular; (c,d) square; (e,f) triangular, apex-to-conduit; (g,h) triangular, apex-to-cavity. The computed pressure is sampled on the centerline of the gap, 6~mm upstream of the wall and 0.5~mm from the end wall, and is smoothed using a 10~$\mu$s moving average. In each panel, $t = 0$ corresponds to the arrival of the first pressure jump at that location. The simulations are initialized with the incident shock between the upstream probe and the wall, so the computed upstream histories begin at the pressure behind the incident shock.}
    \label{fig:HeadOnSimExp}
\end{figure*}

Figure~\ref{fig:HeadOnSimExp} compares the computed and measured pressure histories 6~mm upstream of the wall and at the end wall. The computed histories contain high-frequency oscillations that are not resolved by the transducers, and are therefore presented after a 10~$\mu$s moving average. At the upstream probe, the simulations reproduce the arrival time of the reflected shock, with a difference of less than 5\% in the pressure values for all four geometries. At the end wall, the simulations reproduce the measured sequence: the arrival of the transmitted shock produces an initial jump, the pressure then decreases, and a second jump follows approximately 0.2~ms later, after which the pressure builds up gradually with superimposed reverberations. The mean end-wall pressure during the 0.1~ms that follow the initial jump is 1.4 to 1.7~bar in the simulations and 1.5 to 1.9~bar in the experiments, with the lowest value obtained for the apex-to-cavity wall in both.

The main difference between the two data sets appears at late times, where the computed pressures approach 3.0 to 3.3~bar while the measured pressures approach 3.5~bar. This difference is consistent with the lower incident Mach number of the simulations, $M = 1.4$ compared with $M = 1.45$ in the experiments. The weaker incident shock produces an overpressure of 1.13~bar behind it, compared with approximately 1.2 to 1.3~bar measured at the upstream transducer. It also propagates more slowly, and its ideal reflection from a solid wall gives an overpressure of 3.2~bar. The two-dimensional idealization of a single period of the wall may also contribute. Given that the simulations reproduce the timing and the amplitude of the main features of the pressure histories at both locations, and the differences between the geometries, the agreement is considered sufficient for using them to examine the wave propagation inside the wall.

\begin{figure*}[t]
    \centering
    \includegraphics[width=0.9\textwidth]{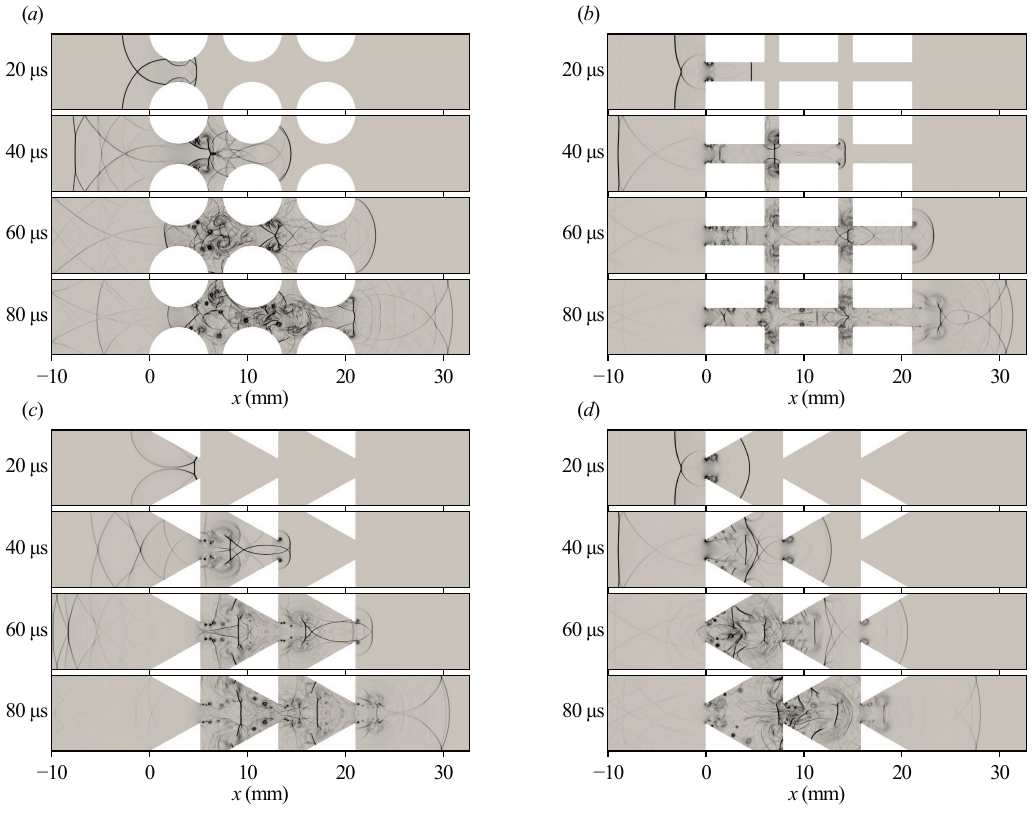}
    \caption{Numerical schlieren images of the shock propagation through one period of the porous wall under head-on impingement, at $t = 20$, 40, 60 and 80~$\mu$s: (a) circular; (b) square; (c) triangular, apex-to-conduit; (d) triangular, apex-to-cavity. The incident shock propagates from left to right. The front and rear faces of the wall are located at $x = 0$ and $x = 21$~mm and the end wall is located at $x=60$~mm. Time is measured from the beginning of the simulation, when the incident shock is located at $x = -5$~mm. The shock reaches the front face of the wall at $t \approx 10~\mu$s.  }
    \label{fig:HeadOnSchlieren}
\end{figure*}

Figure~\ref{fig:HeadOnSchlieren} presents numerical schlieren images of the four geometries at four instants, covering the passage of the shock through the three rows of elements and its emergence from the rear face of the wall. The first difference between the geometries appears at the front face. For the square and apex-to-cavity walls, the incident shock meets flat surfaces that block three quarters of the cross-section. These surfaces reflect it as a nearly planar shock, and only the segment facing the gap enters the wall, with vortices forming at the sharp edges of the entrance (Figure~\ref{fig:HeadOnSchlieren}\textit{(b,d)}, 20~$\mu$s). For the apex-to-conduit wall, the shock meets inclined surfaces that guide the entire front toward the gap. The waves reflected from these surfaces are weak and curved, and the front reaches the first gap as a short stem bounded by them (Figure~\ref{fig:HeadOnSchlieren}\textit{(c)}, 20~$\mu$s), in agreement with the transition to Mach reflection and the weak reflected wave reported for upstream-pointing triangles in \cite{Chaudhuri2013}. The reflected waves then merge into a planar front only farther upstream, which is consistent with the more gradual pressure rise measured upstream of this wall (Figure~\ref{fig:HeadOnSimExp}\textit{(e)}). The circular wall shows an intermediate behavior, with curved reflected waves originating from the upstream surfaces of the cylinders. Since the four walls share the same frontal blockage, the upstream pressure subsequently approaches a similar level for all of them (Figure~\ref{fig:HeadOnSimExp}\textit{(a,c,e,g)}).

Inside the wall, the evolution of the transmitted front depends on how the passage area changes along its path. In the square wall the passage has a constant width, and the front remains planar throughout the wall. It is disturbed only where it crosses the lateral slots separating successive rows, into which it diffracts and where vortices form at the corners (Figure~\ref{fig:HeadOnSchlieren}\textit{(b)}, 40 and 60~$\mu$s). In the apex-to-conduit wall, the front expands abruptly behind the bases of each row and is then collected again by the converging surfaces of the following row. The waves reflected from these surfaces cross on the centerline behind the front, which leaves each gap as a short planar segment (Figure~\ref{fig:HeadOnSchlieren}\textit{(c)}, 40 and 60~$\mu$s). The reversed orientation produces the opposite sequence. The front leaves the gap at the front face directly into a widening passage and expands as a curved front. By the time it reaches the following row it spans the entire passage, so most of it meets the flat bases of that row and is reflected back, and only its central part continues through the next gap (Figure~\ref{fig:HeadOnSchlieren}\textit{(d)}, 40~$\mu$s). This process repeats at each of the three rows, which leads to the maximum attenuation of the transmitted shock in this case. The circular wall combines a gradual contraction and a gradual expansion at each row, and the front diffracts around the downstream side of the cylinders, where the flow separates, and vortices are shed into the region between the rows (Figure~\ref{fig:HeadOnSchlieren}\textit{(a)}, 40 and 60~$\mu$s).

These differences determine the strength and the timing of the front that emerges from the wall. At 60~$\mu$s the transmitted front has left the rear face of the circular, square and apex-to-conduit walls, while for the apex-to-cavity wall it is still inside the last row. At 80~$\mu$s the leading front is located at $x \approx 30$ to 31~mm for the first three walls and at $x \approx 27.5$~mm for the apex-to-cavity wall. In all cases the emerging front is initially curved and flattens through successive reflections from the lateral boundaries as it propagates toward the end wall, forming the unified transmitted front. The delay accumulated by the apex-to-cavity wall persists to the end wall. Its transit time from the upstream transducer to the end wall exceeds those of the other three walls by 11 to 17~$\mu$s in the simulations and by 3 to 11~$\mu$s in the experiments, and its transmitted shock produces the lowest end-wall pressure following the initial jump (Figure~\ref{fig:HeadOnSimExp}\textit{(h)}). The lower transmitted-front speed for this orientation agrees with the observations of \cite{Chaudhuri2013} for downstream-pointing triangles. The two orientations of the triangular elements therefore differ in the way the wall attenuates the incident shock. The apex-to-conduit wall reflects weakly and passes most of the front through its converging passages. The apex-to-cavity wall reflects strongly at its front face and again at every row, and transmits the weakest and most delayed front among the four geometries.

The images also show the flow that follows the transmitted front. In the apex-to-cavity wall, the flow separates at the sharp edges of each gap and enters the widening passage as a jet bounded by vortices, rather than following the diverging surfaces (Figure~\ref{fig:HeadOnSchlieren}\textit{(d)}, 60 and 80~$\mu$s). In the apex-to-conduit wall, the flow approaches the first gap along the converging surfaces without visible separation, and separates at the corners of the bases, downstream of each gap (Figure~\ref{fig:HeadOnSchlieren}\textit{(c)}, 60 and 80~$\mu$s). These observations cover only the first stage of the process, which is short compared with the filling times of the confined volume. They nevertheless bear on the dependence of the pressure build-up on element orientation examined in Section~\ref{sec:Head-onCalibration}: separation at the entrance of a widening passage reduces the effective flow area.

\begin{figure*}[t]
    \centering
    \includegraphics[width=\textwidth]{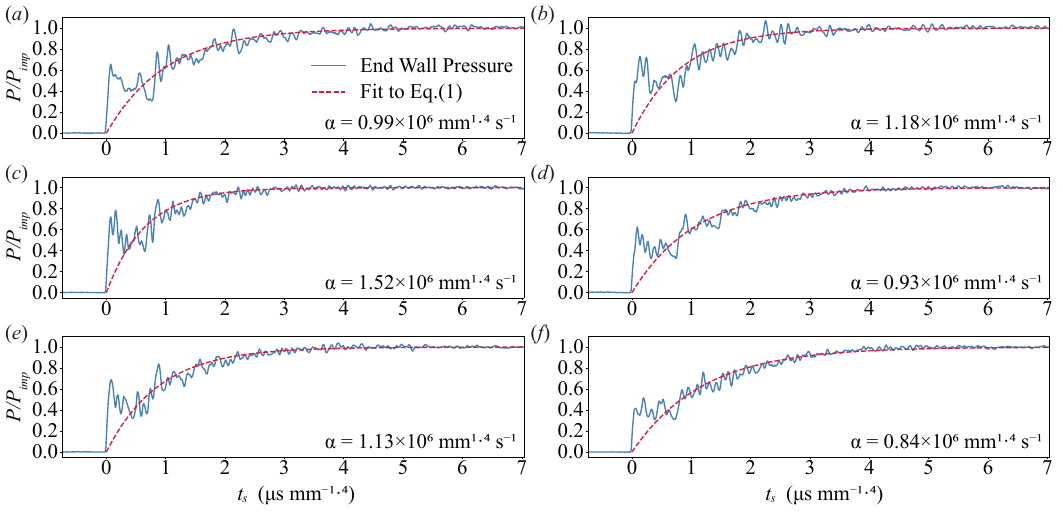}
    \caption[]{Fitting results for a representative experiment of the head-on impingement of a shock wave on porous models of varying internal geometry: (a) circular; (b) square; (c) triangular, apex-to-conduit; (d) triangular, apex-to-cavity; (e) staggered triangular, apex-to-conduit; (f) staggered triangular, apex-to-cavity.}
    \label{Headon_P_tau}
\end{figure*}

\begin{table*}[t]
    \centering
    \caption[]{Lumped parameter $\alpha$ and time constant $\tau_{\mathrm{HO}} = (\mathrm{SOD} + L_{\mathrm{eff}})^{\gamma}/\alpha$ from the head-on experiments, $\mathrm{SOD} = 39$~mm, $M = 1.45$. $n$ is the number of repeats.}
    \label{tab:headon}
    \begin{tabular}{lcccccc}
        \hline
        Internal geometry & $\phi$ & $L_{\mathrm{eff}}$ (mm) & $n$ & $\alpha$ ($10^{6}$~mm$^{\gamma}$\,s$^{-1}$) & $\tau_{\mathrm{HO}}$ ($\mu$s) \\
        \hline
        Circular                               & 0.495 & 10.40 & 5 & $1.004 \pm 0.036$ & $234.4 \pm 8.3$ \\
        Square                                 & 0.357 & 7.50 & 4 & $1.193 \pm 0.020$ & $181.1 \pm 3.0$ \\
        Triangular, apex-to-conduit               & 0.722 & 15.16 & 3 & $1.526 \pm 0.038$ & $175.3 \pm 4.3$ \\
        Triangular, apex-to-cavity                & 0.722 & 15.16 & 5 & $0.923 \pm 0.017$ & $289.8 \pm 5.3$ \\
        Triangular, apex-to-conduit, staggered    & 0.740 & 15.54 & 5 & $1.126 \pm 0.031$ & $239.9 \pm 6.6$ \\
        Triangular, apex-to-cavity, staggered     & 0.740 & 15.54 & 4 & $0.839 \pm 0.004$ & $322.0 \pm 1.5$ \\
        \hline
    \end{tabular}
\end{table*}
\subsubsection{Pressure response and model calibration}
\label{sec:Head-onCalibration}

The head-on experiments were used to calibrate Eq.~\eqref{eq:Ram_Model} for each of the six internal geometries, at a stand-off distance of $39$~mm and an incident Mach number of $M = 1.45$. Figure~\ref{Headon_P_tau} presents the normalized pressure histories on the end wall as functions of the scaled time $t_s$, together with the curves obtained from Eq.~\eqref{eq:Ram_Model} with the fitted values of $\alpha$. All six geometries reproduce the behavior reported in~\cite{ram2013simple}: The transmitted shock arrives at the end wall and produces an initial pressure jump, followed by progressively weaker jumps as the trapped shock reverberates between the back face of the model and the end wall, superimposed on a monotonic build-up driven by the filtration flow through the model. The fitted values of $\alpha$, collected in Table~\ref{tab:headon}, span $0.839$ to $1.526\times10^{6}$~mm$^{\gamma}$\,s$^{-1}$ and bracket the $1.139\times10^{6}$~mm$^{\gamma}$\,s$^{-1}$ reported in~\cite{ram2013simple} for the 10-PPI SiC foam sample.

The orientation of the triangular elements has the largest effect on $\alpha$. Within each arrangement, the two orientations share the same void fraction and frontal blockage, and differ only in whether the through-passage converges or diverges in the direction of the transmitted flow. Yet with the apex facing the conduit, $\alpha$ is larger by a factor of $1.65$ in the aligned arrangement and $1.34$ in the staggered one, and the confined volume fills correspondingly faster, with $\tau_{\mathrm{HO}}$ decreasing from $290$ to $175~\mu$s and from $322$ to $240~\mu$s, respectively. With the apex facing the conduit, the passage between adjacent triangular elements narrows in the direction of flow, reaching its smallest gap between their bases. With the apex facing the cavity, the flow encounters this gap first, followed by a widening passage. Because the wall contains three rows of elements, the flow experiences successive contractions and expansions as it passes through the wall. The head-on simulations show that the early separation patterns of the two aligned triangular walls differ. With the apex facing the cavity, the flow separates at the entrance to each widening passage; with the apex facing the conduit, separation occurs downstream of the minimum gap (Figure~\ref{fig:HeadOnSchlieren}). These observations are consistent with a smaller effective flow area and a slower pressure build-up for the apex-to-cavity orientation, although they do not quantify the effective resistance over the full filling process. In the staggered arrangement, the offset rows also require the flow to turn between successive rows. These additional changes in flow direction may reduce the relative influence of element orientation, which could explain the smaller difference between the two orientations in this arrangement. The staggered walls were not simulated, so this role of flow turning remains an interpretation of the measurements. The two apex-to-cavity walls exhibit the longest characteristic filling times, a result that becomes relevant under parallel loading (Section~\ref{sec:parallel}).

The smooth exponential model does not capture the finite initial pressure jump or the subsequent reverberations. Ram and Sadot~\cite{ram2013simple} identify this as a limitation of a model derived under the assumption of isentropic filling, and note that both low PPI and stronger incident shocks reduce its accuracy. The present walls are more open than their $10$--$30$~PPI foams, and the ordered grid samples in that study were validated only over $1.17 \le M \le 1.22$~\cite{ram2013simple} compared with $M = 1.45$ here. Because Eq.~\eqref{eq:Ram_Model} starts from zero while the measured end-wall pressure jumps to a large fraction of $P_{imp}$ on arrival of the transmitted shock, the fitted $\alpha$ represents an effective response rate that incorporates both the initial jump and the subsequent equilibration. The six head-on coefficients provide the calibration used in Section~\ref{sec:parallel} to test whether the same response timescale describes loading that progresses along the wall.

Taken together, the wave patterns in Figure~\ref{fig:HeadOnSchlieren} and the coefficients of Table~\ref{tab:headon} help interpret the trends reported in the following sections. Under head-on impingement, the apex-to-cavity wall transmits the weakest front and presents an abrupt contraction to the flow at every row. Of the four aligned geometries, it has the lowest build-up coefficient and the longest filling time. The apex-to-conduit wall is at the other extreme, with the highest coefficient and the shortest filling time (Table~\ref{tab:headon}). The same mechanisms carry over to parallel propagation (Section~\ref{sec:parallel}). There the loading progresses along the wall, but each wall still presents the same surface to the conduit, and the disturbances transmitted toward the cavity meet the same passage shape. As Section~\ref{sec:parallel} shows, on the conduit side, at the two larger cavity heights, the square and apex-to-cavity walls, which present flat surfaces to the conduit and reflect the incident shock most strongly, retain higher conduit-front velocities and peak pressures than the apex-to-conduit wall, which admits the wave more readily into its passages. On the cavity side, the weak and delayed transmission through the apex-to-cavity wall is consistent with the low initial pressure jump, the longer subsequent build-up, and the persistent arrival lag found for this orientation.

\subsection{Parallel Propagation}
\label{sec:parallel}
In this configuration, the shock propagates along the same six porous walls tested under head-on impingement, loading each wall progressively along its length. Three cavity heights were tested, $H_1$, $H_2$ and $H_3$ ($9$, $39$ and $69$~mm, respectively) as seen in Figure~\ref{fig:Experimental}\textit{(b)}. The intermediate height matches the $39$~mm stand-off distance used in the head-on experiments, allowing comparison at the same wall-to-boundary separation. All of the parallel propagation experiments were performed with an incident shock wave Mach number of 1.45 at the entrance to the test section and were repeated a minimum of 12 times per configuration.

\begin{figure*}[t]
    \centering
    \includegraphics[width=\textwidth]{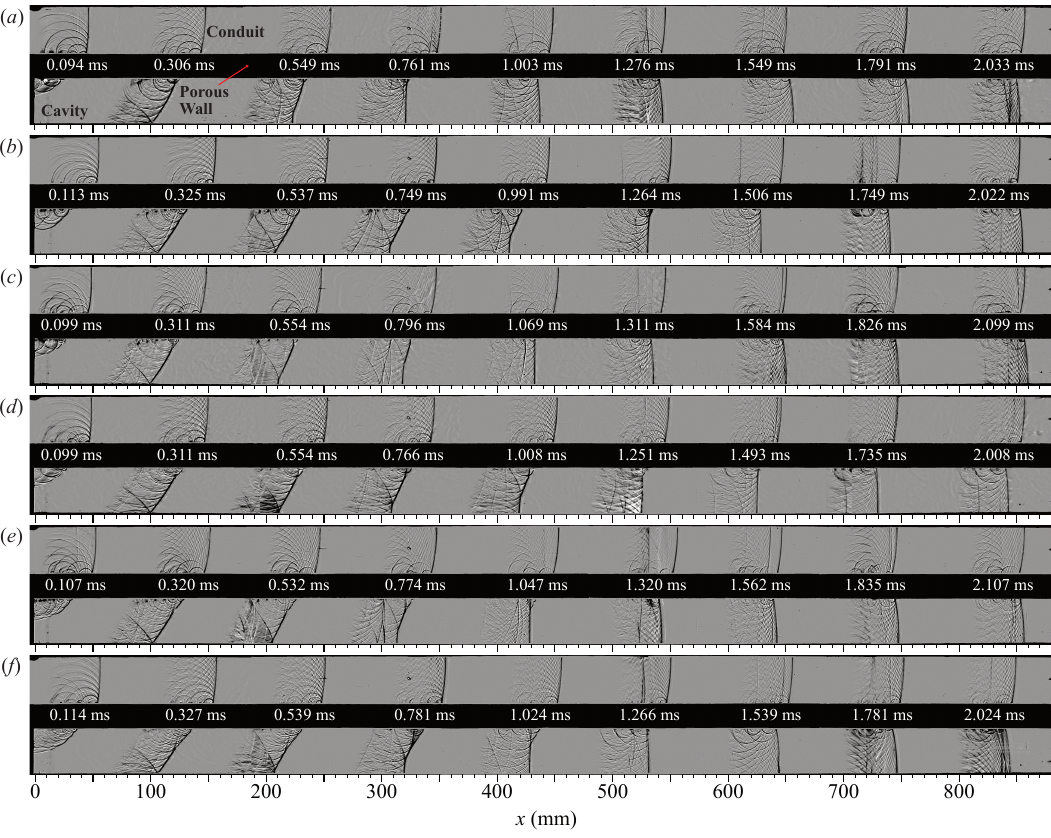}
    \caption[]{ Composite schlieren images showing shock-front evolution in the conduit (above the porous wall) and cavity (below) at a cavity height of $39$~mm ($H_2$). Each panel combines images from separate recordings, showing the fronts at conduit-shock positions of approximately $50$ to $850$~mm in $100$~mm increments. Times are measured from the conduit shock's arrival at the cavity entrance ($x = 0$). Porous wall geometries: (a) circular cylinders; (b) squares; (c) aligned triangles, apex facing the conduit; (d) aligned triangles, apex facing the cavity; (e) staggered triangles, apex facing the conduit; and (f) staggered triangles, apex facing the cavity. }
    \label{fig:schlieren}
\end{figure*}

\subsubsection{Schlieren visualization of shock-front evolution}
Figure~\ref{fig:schlieren} compares the evolution of the conduit and cavity fronts for the six porous wall geometries at the intermediate cavity height, $H_2$. Each composite shows successive stages selected at approximately matching conduit-front positions, allowing the cavity-front shape and its position relative to the conduit front to be compared across geometries. The corresponding times are indicated separately for each panel.

In the conduit, as the incoming shock wave emerges into the test section, it experiences an abrupt area expansion while simultaneously coming into contact with the surface of the porous wall. Following this initial contact, the incoming shock is partially transmitted through the wall and partially reflected from it toward the conduit wall, as seen on the left side of Figure~\ref{fig:schlieren}. The reflected waves then reflect from the conduit wall back toward the porous wall. This sequence repeats as the conduit shock propagates downstream. These repeated reflections eventually form secondary shock waves that follow the main incident front, as seen in Figure~\ref{fig:schlieren} when the conduit front is at approximately 550~mm. The secondary shock waves eventually catch up with the main front, strengthening it and, in some cases, creating complex Mach reflection patterns.
\begin{figure*}[t]
    \centering
    \includegraphics[width=0.9\textwidth]{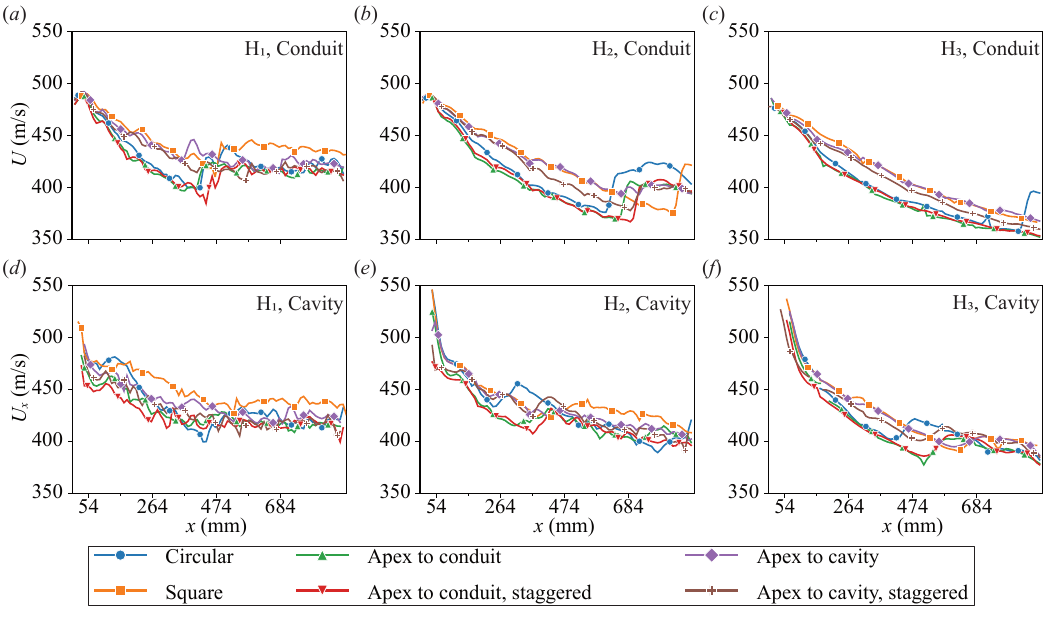}
    \caption[]{Streamwise evolution of shock-front velocities for the six porous wall geometries. Panels (a-c) show the conduit-front velocity, and panels (d-f) show the cavity-front axial velocity evaluated at mid-height. Columns correspond to cavity heights $H_1 = 9$~mm, $H_2 = 39$~mm, and $H_3 = 69$~mm, respectively. Curves represent averages over the available repeated schlieren measurements. Colors and marker shapes identify the wall geometries. The streamwise coordinate $x$ is measured from the cavity entrance.}
    \label{Fig:FrontVel}
\end{figure*}

For all geometries, the shock evolution inside the cavity can be described as a three-stage process, consisting of initial expansion, reflection-driven evolution, and near-normal propagation. Initially, after the passage of the shock wave in the conduit, transmitted shock fronts emerge from the back face of the porous wall and coalesce into a unified front that expands both laterally along the wall and vertically toward the cavity outer wall. The second stage begins when the expanding front reaches the outer wall and reflects from it as a regular reflection. The reflection divides the front into two segments that meet at a reflection point moving along the outer wall: a curved incident front, which spans the cavity between the reflection point and the porous wall and travels downstream, and a reflected front, which propagates back from the outer wall toward the porous wall. As the reflected front approaches the porous wall, it interacts with the flow that continues to emerge through the wall and gains momentum in the downstream direction. The inclination of the incident front, in turn, changes through its interaction with the porous-wall geometry. As the angle between the incident front and the outer wall increases at the reflection point, a Mach stem appears at the wall, marking the transition from regular to Mach reflection. The Mach-reflection structure consists of an incident shock, a reflected shock and a Mach stem meeting at a triple point. As the pattern continues to propagate downstream, the Mach stem grows and the triple point moves away from the outer wall toward the porous wall. The third stage begins once the triple point arrives at the porous wall. From there on, the leading front remains nearly normal to the direction of propagation as it continues downstream. It still interacts with the flow entering through the porous wall and with the reflected shocks behind it.

The streamwise extent of this evolution depends on wall geometry. For the circular wall, a near-normal cavity front is evident in the composite by the window at a conduit-front position of approximately $350$~mm (Figure~\ref{fig:schlieren}\textit{(a)}). The square and aligned apex-to-cavity walls reach a comparable stage around the $550$~mm window (Figure~\ref{fig:schlieren}\textit{(b,d)}). Among the aligned triangular walls, the apex-to-conduit arrangement reaches this stage earlier in the sequence (Figure~\ref{fig:schlieren}\textit{(c,d)}). The staggered arrangements reach it at similar conduit-front positions, although their intermediate front shapes differ (Figure~\ref{fig:schlieren}\textit{(e,f)}).

Cavity height also affects the spatial extent of the front evolution. The schlieren recordings at $H_1$ and $H_3$ show that, for a given wall geometry, the inclined cavity front generally persists farther downstream as the cavity height increases. In the shallow cavity, the expanding front reaches the outer wall close to the cavity entrance, and the subsequent regular-reflection and Mach-reflection stages occur over a short streamwise distance. In the taller cavity, the front traverses a greater transverse distance, retaining its curved and inclined form farther downstream. The Mach-reflection stage also extends over a longer streamwise distance.

The image sequences also show changes in the relative positions of the fronts in the two volumes. Early in the sequence, the cavity front near the outer wall trails the conduit shock in every geometry. This separation generally decreases downstream, but its subsequent evolution differs between geometries. For the circular wall, the cavity front becomes visibly advanced relative to the conduit front in the later stages before the fronts approach one another again near the end of the sequence. Thus, the initial cavity-front lag does not persist throughout the test section in every case.

The wave pattern behind the leading fronts becomes increasingly complex downstream, with closely spaced fronts visible in both volumes. These structures provide context for the secondary features in the pressure histories discussed below. The following subsection quantifies the front velocities, the triple-point trajectories, and the cavity-to-conduit arrival lag.

\begin{figure}[t]
    \centering
    \includegraphics[width=0.50\textwidth]{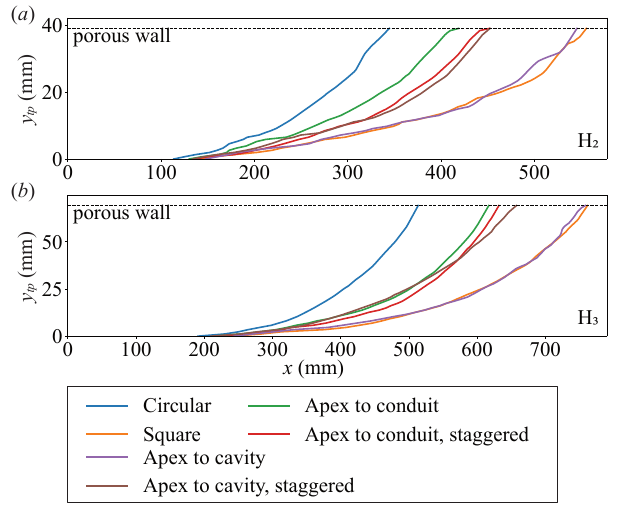}
    \caption[]{Triple-point trajectories in the cavity for the six porous wall geometries at (a) $H_2 = 39$~mm and (b) $H_3 = 69$~mm. The transverse distance is measured from the cavity outer wall, and $x$ denotes the streamwise position of the triple point measured from the cavity entrance. Curves represent the mean streamwise position at each transverse distance over the repeated recordings. Dashed lines indicate the porous-wall location. }
    \label{Fig:TriplePoint}
\end{figure}

\subsubsection{Shock-front kinematics}
\label{sec:ShockKinematics}

The schlieren recordings were used to quantify the downstream progression of the fronts in both volumes and the development of the Mach-reflection structure in the cavity. The analysis considers the conduit-front velocity, the axial velocity of the cavity front at mid-height, the triple-point trajectories, and the cavity-to-conduit arrival-time difference.

Figure~\ref{Fig:FrontVel} compares the front velocities for the six porous wall geometries at the three cavity heights. The conduit shock initially decelerates in every case. Farther downstream, the evolution depends on both wall geometry and cavity height. At $H_1$, the initial deceleration is interrupted by local velocity recoveries, after which the velocities remain approximately within $410$ to $440$~m/s over much of the downstream region (Figure~\ref{Fig:FrontVel}\textit{(a)}). At $H_2$, the deceleration extends farther downstream, with pronounced recoveries occurring at different locations for several geometries (Figure~\ref{Fig:FrontVel}\textit{(b)}). At $H_3$, most geometries continue to decelerate over most of the observed region, approaching approximately $350$ to $370$~m/s near its downstream end. The circular geometry exhibits a late recovery (Figure~\ref{Fig:FrontVel}\textit{(c)}).

The abrupt conduit-velocity recoveries are consistent with reinforcement of the leading shock by the secondary shocks observed in the schlieren recordings. As a secondary shock catches the leading front, their interaction can produce a stronger outgoing shock and interrupt the preceding deceleration. This provides a physical explanation for the local velocity increases, subject to correspondence between the observed merger and recovery locations for each case. The shift of the recoveries farther downstream with increasing cavity height is in turn consistent with a later interaction between the secondary and leading fronts. The pressure histories discussed below provide complementary evidence through the evolution of the leading jump and secondary pressure steps.

Wall geometry also affects the conduit-front velocity before these recoveries. At $H_2$ and $H_3$, the square and aligned apex-to-cavity walls generally retain higher conduit velocities than the two apex-to-conduit arrangements over the intermediate part of the test section (Figure~\ref{Fig:FrontVel}\textit{(b,c)}). The subsequent recoveries modify this ordering. Geometry therefore affects both the initial deceleration and the downstream evolution associated with interactions between the leading and secondary waves.

The cavity front also exhibits a rapid initial reduction in axial velocity, followed by a more gradual downstream evolution with local increases and decreases (Figure~\ref{Fig:FrontVel}\textit{(d, e, f)}). Its downstream velocity generally decreases with increasing cavity height, although the separation between heights is less pronounced than in the conduit. At $H_2$ and $H_3$, the cavity front at mid-height advances faster than the conduit front over substantial portions of the test section. The cavity front can therefore reduce its initial lag even while it decelerates, provided that the conduit front advances more slowly.

This velocity comparison must also account for the changing cavity-front shape. During the inclined stage, the axial velocity at mid-height differs from the local propagation velocity normal to the front. A higher axial velocity therefore does not by itself imply a stronger cavity shock. Following the evolution toward a near-normal front, the axial and normal directions become approximately aligned. Any remaining axial velocity advantage can then no longer be attributed primarily to front inclination. Since the velocity is evaluated at mid-height, its connection to the outer-wall arrival measurements becomes more direct once the front is nearly normal across the cavity.
\begin{figure*}[t]
    \centering

    \includegraphics[width=0.88\textwidth]{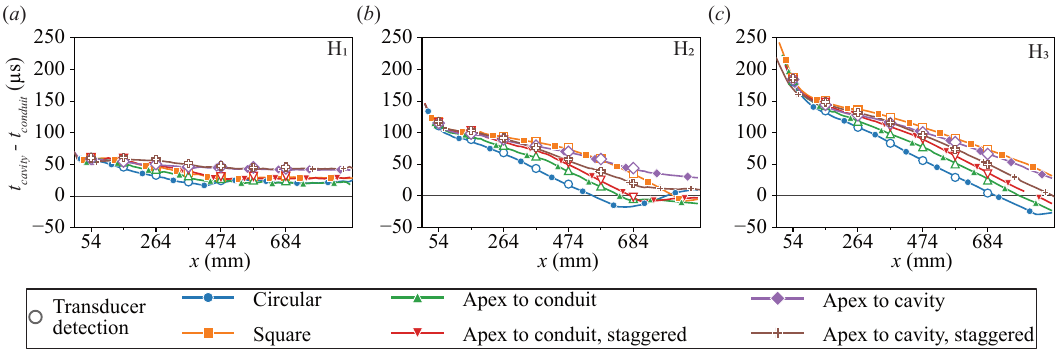}
    \caption[]{Cavity-to-conduit front arrival lag for the six porous wall geometries at (a) $H_1 = 9$~mm, (b) $H_2 = 39$~mm, and (c) $H_3 = 69$~mm. Lines and filled markers show averages from schlieren measurements evaluated near the cavity outer wall. Open markers show averages from paired pressure measurements. Positive lag indicates that the cavity front arrives after the conduit front, while negative lag indicates the reverse arrival order. The horizontal line marks simultaneous arrival. The streamwise coordinate $x$ is measured from the cavity entrance.}
    \label{Fig:FrontLag}
\end{figure*}

The triple-point trajectories in Figure~\ref{Fig:TriplePoint} quantify the development of the Mach-reflection structure at $H_2$ and $H_3$. The transverse coordinate is measured from the cavity outer wall, and the horizontal coordinate denotes the streamwise position of the triple point. Progression toward the porous wall describes the growth of the Mach stem across the cavity. A steeper trajectory indicates greater transverse advance per unit downstream distance, rather than directly specifying a larger transverse velocity. The $H_1$ case is excluded because the triple point could not be identified with sufficient confidence throughout its evolution.

At both heights, the circular geometry reaches the full-span state over the shortest streamwise distance, whereas the square and aligned apex-to-cavity geometries reach the porous wall farthest downstream. The apex-to-conduit arrangements and the staggered apex-to-cavity geometry form an intermediate group. These differences show that wall geometry affects the persistence of the inclined incident-front segment and the streamwise extent of the reflection-driven evolution. They are consistent with geometry-dependent differences in the disturbances and flow emerging through the porous wall, which influence the cavity-front shape and its interaction with the outer wall.

Increasing the cavity height from $H_2$ to $H_3$ shifts the trajectories downstream and extends the region over which the Mach stem grows (Figure~\ref{Fig:TriplePoint}\textit{(a,b)}). The greater transverse distance provides a geometric contribution to this longer evolution. The broadly preserved ordering of the geometries suggests that the same reflection sequence remains relevant at both heights. Nevertheless, the trajectories do not by themselves establish that the height dependence is a simple geometric rescaling. Completion of the triple-point crossing marks the development of a front spanning the cavity in a near-normal form, rather than the attainment of a constant propagation velocity.

Figure~\ref{Fig:FrontLag} presents the difference between the arrival times of the cavity and conduit fronts at the same streamwise position. Positive values indicate that the cavity front lags behind the conduit front, while negative values indicate that it has overtaken the conduit front and is in the lead. This arrival-time difference is compared with the paired pressure measurements on the cavity and conduit walls.

The initial arrival lag depends strongly on cavity height, with values of approximately $60$, $115$, and $185~\mu\mathrm{s}$ at $H_1$, $H_2$, and $H_3$, respectively (Figure~\ref{Fig:FrontLag}\textit{(a,b,c)}). Differences in lag between geometries are small initially, becoming more pronounced downstream. At $H_1$, all geometries show a small decrease in the lag followed by an approximately constant value (Figure~\ref{Fig:FrontLag}\textit{(a)}). At this cavity height, both apex-to-cavity configurations retain the largest lag over the test section. The comparatively weak downstream variation in lag is consistent with the fronts approaching similar axial velocities once the cavity front becomes nearly normal. At both $H_2$ and $H_3$, the lag initially decreases steeply and then declines more gradually, with the downstream lag varying with the internal geometry of the porous wall (Figure~\ref{Fig:FrontLag}\textit{(b,c)}). The ordering of the geometries is largely preserved downstream, except at $H_2$, where it is disrupted as the cavity front overtakes the conduit front. At $H_2$, the cavity front overtakes the conduit front in all geometries except for the two apex-to-cavity arrangements. This occurs first for the circular geometry, then for the two apex-to-conduit configurations, and finally for the square geometry. A similar process occurs at $H_3$, but is cut short by the end of the test section.

The square and aligned apex-to-cavity geometries share a relatively persistent arrival lag, higher intermediate conduit velocities, and later completion of the triple-point crossing at $H_2$ and $H_3$. The two diverge at $H_2$, where the lag of the square geometry begins to fall rapidly at approximately $x = 680$~mm, so that the square-wall cavity front takes the lead somewhere between $810$ and $820$~mm, while the aligned apex-to-cavity wall retains its lag. This divergence is consistent with the earlier acceleration of the conduit front in the aligned apex-to-cavity case. Reinforcement of the leading shock as the secondary shock catches it could account for this acceleration and help the conduit front retain its lead.

The circular geometry provides the clearest example of how the velocity and reflection results combine to determine the arrival lag. At $H_2$, its relatively early evolution toward a near-normal cavity front and the cavity front's velocity advantage over substantial downstream intervals accompany a progressive reduction in lag. The cavity front subsequently arrives before the conduit front over part of the test section. Farther downstream, the recovery of conduit front velocity is consistent with the lag turning upward and eventually becoming positive again (Figure~\ref{Fig:FrontLag}\textit{(b)}). The return to positive lag occurs after the conduit regains a velocity advantage because arrival lag accumulates the preceding differences in travel time. At $H_3$, the circular geometry begins with a larger cavity front lag, and the reversal of arrival order occurs farther downstream (Figure~\ref{Fig:FrontLag}\textit{(c)}). The late conduit-velocity recovery does not restore a positive lag within the observed domain. The pressure-derived arrival differences broadly follow the wall-adjacent optical trends at the measurement stations, supporting the observed reduction and, in some cases, reversal of the cavity-front lag.

Together, the measurements show that relative front arrival is governed by the initial delay, the evolution of cavity front shape, and subsequent velocity changes in both volumes. Wall geometry affects the reflection-driven evolution as well as the relative downstream progression, while cavity height changes the initial lag and the spatial extent over which the fronts evolve. The pressure histories examined next establish how these differences in wave evolution are reflected in the loading of the conduit and cavity.

\subsubsection{Pressure histories}
\label{sec:Pressure}
\begin{figure*}[t]
    \centering
    \includegraphics[width=\textwidth]{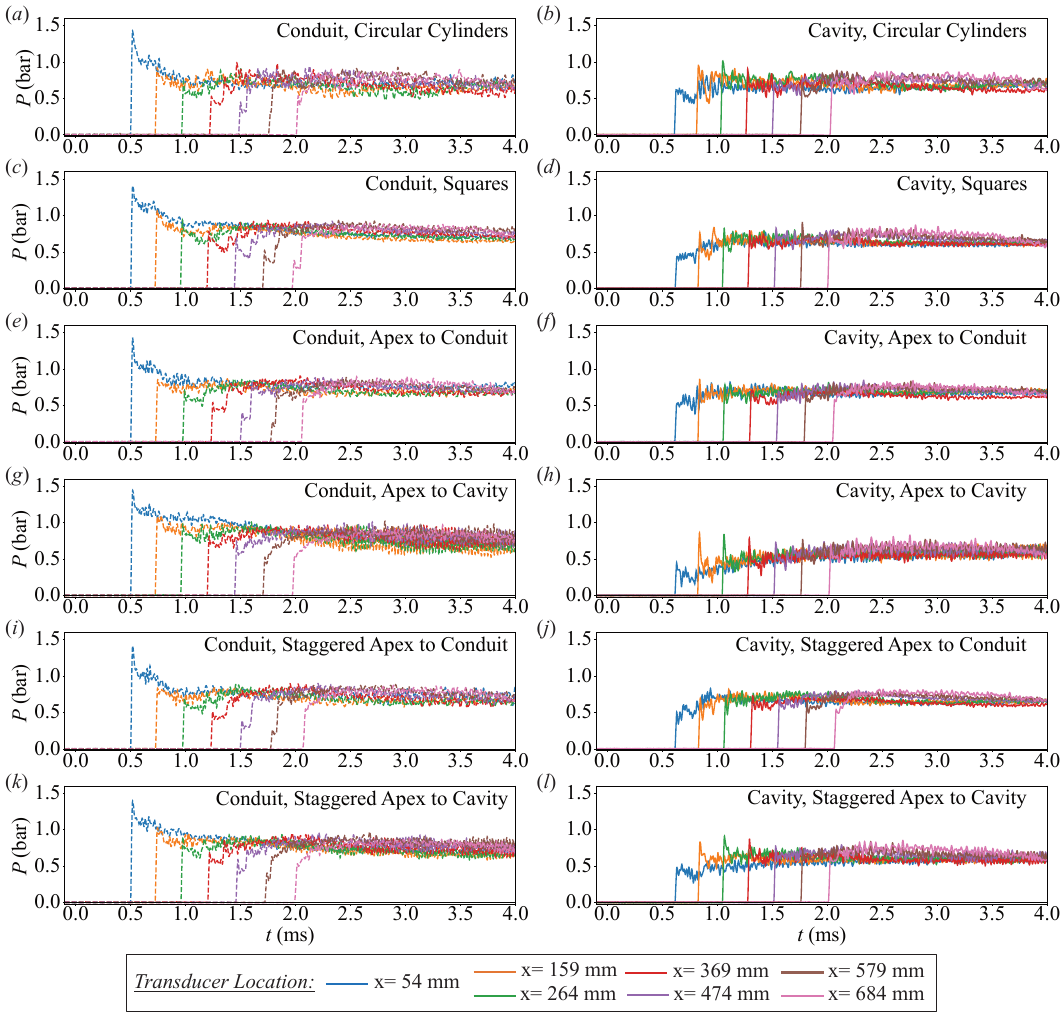}
    \caption[]{Pressure histories on the conduit and cavity sides of the porous wall for the six internal geometries at a cavity height of $39$~mm ($H_2$): (a,b) circular; (c,d) square; (e,f) triangular, apex-to-conduit; (g,h) triangular, apex-to-cavity; (i,j) staggered triangular, apex-to-conduit; (k,l) staggered triangular, apex-to-cavity.}
    \label{Parallel_H2}
\end{figure*}
\begin{figure*}[t]
    \centering
    \includegraphics[width=\textwidth]{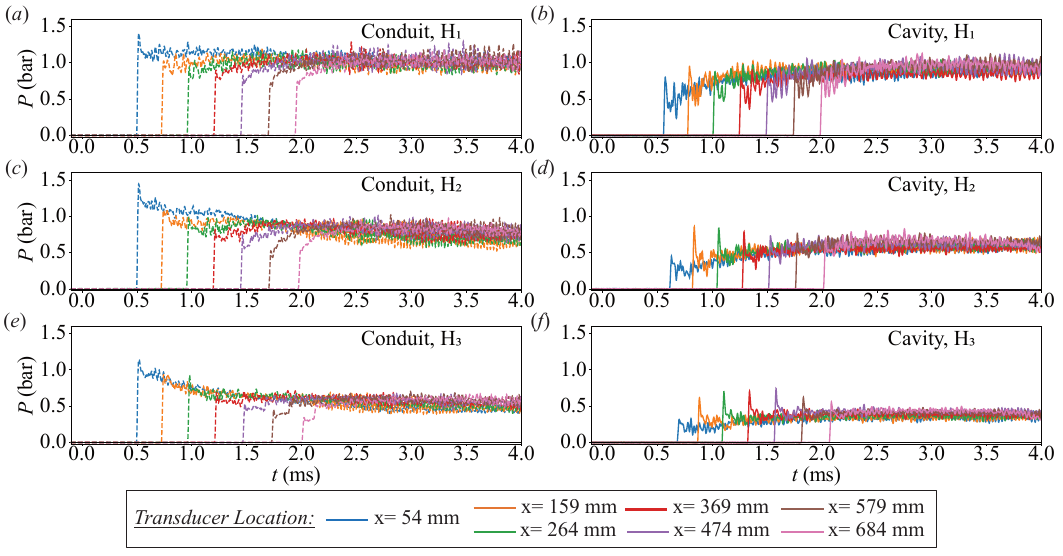}
    \caption[]{Pressure histories for the aligned apex-to-cavity wall at the three cavity heights. Conduit records (a,c,e) and cavity records (b,d,f) for $H_1$, $H_2$ and $H_3$, respectively.}
    \label{Tcavity_H}
\end{figure*}

The pressure histories characterize the loading associated with front arrival and its subsequent evolution in the conduit and cavity. Figure~\ref{Parallel_H2} compares the pressure histories of the six porous wall geometries at $H_2$ in both the conduit (Figure~\ref{Parallel_H2}\textit{(a,c,e,g,i,k)}) and cavity (Figure~\ref{Parallel_H2}\textit{(b,d,f,h,j,l)}). These records complement the front kinematics by distinguishing the initial pressure jump from later pressure changes at each measurement station.

In the conduit records, the initial sharp pressure jump is followed by a pressure decrease toward a slowly varying late-time level. At each station after the first, the pressure jump rises only to the pressure level prevailing at the preceding station at that time. In several downstream records, the pressure jump reaches only part of that level, with a secondary jump providing the remaining rise (Figure~\ref{Parallel_H2}\textit{(a,c,e,i,k)}). These secondary pressure jumps are consistent with the passage of the secondary shock front identified in the schlieren recordings (Figure~\ref{fig:schlieren}). Shortly after the secondary shock wave forms, the initial pressure jump is large and the interval between the two jumps is long. As the secondary front closes on the leading conduit front, this interval shortens from station to station, and once the fronts merge the record returns to a single jump. The same development is evident in the front velocities (Figure~\ref{Fig:FrontVel}\textit{(c)}). For walls whose velocity recovery occurs beyond the last pressure transducer, such as the square wall, this sequence is not completed within the pressure records, whereas for walls in which the fronts merge early, such as the circular wall, it is completed first. The development of the secondary step therefore depends on wall geometry. It is comparatively weak for the aligned apex-to-cavity wall (Figure~\ref{Parallel_H2}\textit{(g)}). The temporal decay following each jump is distinct from the downstream attenuation of the leading-front peak, which is quantified in the following subsection.

The cavity histories are also similar across the geometries; the main difference occurs after the initial pressure jump at station~1. For most of the geometries, the initial disturbance brings the pressure close to its late-time level and is followed by a short build-up (Figure~\ref{Parallel_H2}\textit{(b,d,f,j)}). At station~2, the jump reaches the late-time level almost immediately, and the subsequent oscillations quickly converge to that level. The apex-to-cavity wall configurations, both aligned and staggered, behave differently (Figure~\ref{Parallel_H2}\textit{(h,l)}). In both, the initial jump at station~1 is followed by a pressure drop and a more prolonged recovery. At station~2, the jump overshoots the late-time level, drops sharply to the pressure level at station~1, and then rises gradually toward the late-time level. This sequence repeats at the subsequent stations and subsides by about the fifth station. In comparison to the staggered configuration, the aligned apex-to-cavity wall presents more pronounced overshoots and pressure drops. Thus, changing the arrangement of elements with the same apex orientation alters the pressure history as well as the front evolution.

At $H_2$, where the cavity height matches the stand-off distance of the head-on experiments, the gradual pressure rise at the first cavity station for the aligned apex-to-cavity wall resembles the end-wall pressure build-up under head-on loading. In both configurations, this rise is consistent with continued flow through the porous wall after the initial disturbance. Farther downstream in the parallel configuration, however, the local pressure can already exceed its late-time level at front arrival, as seen in Figure~\ref{Parallel_H2}\textit{(h)}.

Figure~\ref{Tcavity_H} shows how cavity height changes the relative prominence of the key features in pressure evolution in both conduit and cavity for the aligned apex-to-cavity wall. These features include the cavity's initial pressure peak and subsequent build-up and the duration of the secondary pressure step in the conduit. In the cavity, all three heights share the behavior described above: a relatively low initial pressure jump at the first station, followed by an overshoot at the next stations. This behavior becomes more pronounced as the cavity height increases. At $H_1$, the overshoot observed at several downstream stations at $H_2$ is confined to station~2, and station~3 reaches the late-time level rapidly (Figure~\ref{Tcavity_H}\textit{(b)}). At $H_3$, the increase in cavity height amplifies the overall behavior. The initial pressure jump becomes very weak, reaching only $0.35$~bar, and the subsequent overshoot at the later stations remains evident through the last measurement station (Figure~\ref{Tcavity_H}\textit{(f)}). In this case, the overshoot peaks are followed predominantly by relaxation, as the pressure drop that follows brings them to the late-time pressure level.

This downstream change is consistent with the reflection-driven evolution observed in the schlieren recordings. The pressure sensors on the cavity outer wall encounter the developing reflected-wave structure and subsequently the growing Mach stem. Changes in this structure can alter the local pressure peak even while the front's axial velocity decreases. Continued flow through the porous wall contributes to the later pressure evolution, but its effect is superimposed on the initial wave loading and subsequent wave interactions. The sustained rise at $H_1$ and the downstream overshoot at $H_3$ therefore indicate different balances between the initial disturbance and the later pressure response, rather than a uniform effect of cavity height at every station.

The conduit pressure history also depends on cavity height. At $H_1$, the pressure at the first station falls only slightly after the initial jump and soon settles near its late-time level. At subsequent stations, the initial jump is followed by a gradual rise (Figure~\ref{Tcavity_H}\textit{(a)}). At both $H_2$ and $H_3$, the first-station pressure relaxes from a more pronounced initial peak. The clearest difference between these two heights appears farther downstream. The secondary pressure step is weak at $H_2$, whereas at $H_3$ an intermediate pressure level separates the initial and secondary jumps and remains discernible in the record at the last station (Figure~\ref{Tcavity_H}\textit{(c,e)}).

The intermediate level at $H_3$ indicates that the leading front accounts for only part of the local pressure rise. A second disturbance supplies a further pressure increment when it reaches the same station. This is consistent with the secondary front identified in the schlieren recordings. The persistence of two distinct arrivals through the last station suggests that the two fronts remain separated over the instrumented length at $H_3$. At $H_2$, the secondary step is weak enough that its disappearance from a pressure record alone gives no clear indication of where the fronts merge.

\begin{figure*}[t]
    \centering
    \includegraphics[width=\textwidth]{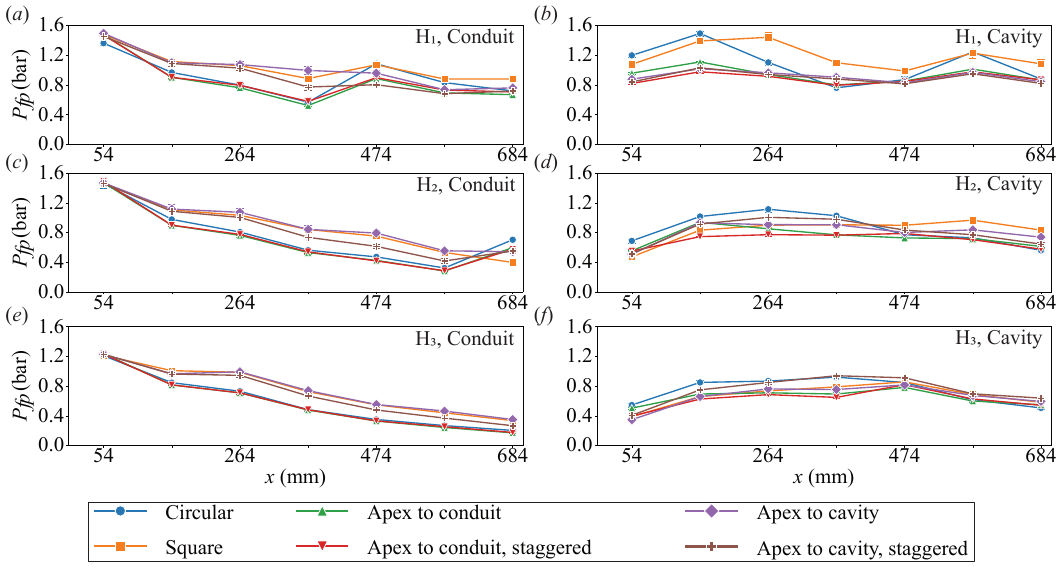}
    \caption{First-rise peak pressure $P_{fp}$ at the seven measurement stations for the six porous wall geometries. Panels \textit{(a,c,e)} show the conduit and panels \textit{(b,d,f)} show the cavity. Rows correspond to $H_1$, $H_2$, and $H_3$ from top to bottom. Markers show the mean over repeated records, and error bars span the minimum and maximum values.}
    \label{fig:firstRisePeak}
\end{figure*}

The downstream cavity records for the aligned apex-to-cavity wall illustrate why the peak pressure and the late-time pressure level must be considered separately. A station can experience a pronounced overshoot at front arrival even when its pressure subsequently relaxes to a level close to that at the preceding station. The following subsection compares the first-rise peak pressure along each volume and examines how much of the late-time cavity pressure is reached on the first rise at station~1.

\subsubsection{Pressure analysis}
\label{sec:P_analysis}

\begin{figure*}[t]
    \centering

    \includegraphics[width=0.8\textwidth]{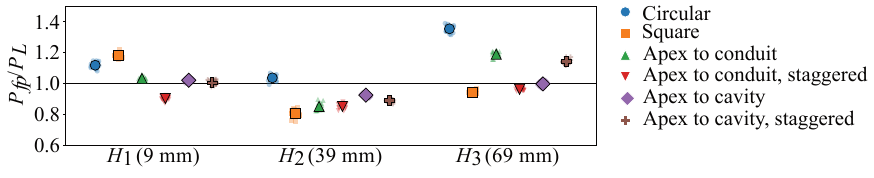}
    \caption{Ratio of the first-rise peak pressure $P_{fp}$ to the late-time pressure $P_L$ at cavity station~1 for the six porous wall geometries at $H_1$, $H_2$, and $H_3$. Both pressures are measured relative to the local pre-arrival baseline, and $P_L$ is averaged from $2.5$ to $3.5$~ms after front arrival. Light symbols show individual shots, and outlined symbols show the mean ratio for each geometry and height.}
    \label{fig:firstRiseFraction}
\end{figure*}

To quantify how the loading at front arrival varies along the porous wall, we compare the first-rise peak pressure $P_{fp}$ at the seven measurement stations in the conduit and cavity (Figure~\ref{fig:firstRisePeak}). All six geometries exhibit the same initial trend at all three cavity heights. From station~1 to station~2, $P_{fp}$ decreases in the conduit and increases in the cavity. For the aligned apex-to-cavity wall at $H_3$, for example, the conduit peak falls from $1.23$ to $0.97$~bar over this interval, while the cavity peak rises from $0.35$ to $0.66$~bar (Figure~\ref{fig:firstRisePeak}\textit{(e,f)}).

Farther downstream, the conduit peak pressures reflect the geometry dependence observed in the front kinematics. The aligned and staggered apex-to-conduit walls follow closely similar pressure curves at all three heights. At $H_2$ and $H_3$, the square and aligned apex-to-cavity walls form a second pair and retain higher peak pressures over the intermediate stations (Figure~\ref{fig:firstRisePeak}\textit{(c,e)}). The separation is particularly clear at station~5 for $H_3$, where the square and aligned apex-to-cavity walls both give approximately $0.55$~bar, compared with $0.33$~bar for both apex-to-conduit arrangements. The same grouping appears in the schlieren measurements. The square and aligned apex-to-cavity walls retain higher conduit-front velocities over the intermediate region and a more persistent cavity-front lag, while the two apex-to-conduit arrangements show similar velocity and lag histories (Figure~\ref{Fig:FrontVel}\textit{(b,c)} and Figure~\ref{Fig:FrontLag}\textit{(b,c)}). The pressure measurements therefore show that the higher conduit-front velocities of the square and aligned apex-to-cavity walls are accompanied by larger pressure jumps at front arrival.

The circular wall exhibits the largest local recovery in conduit peak pressure at both $H_1$ and $H_2$. At $H_1$, $P_{fp}$ rises from $0.57$ to $1.08$~bar between stations~4 and~5. At $H_2$, the recovery occurs farther downstream, with $P_{fp}$ rising from $0.32$ to $0.70$~bar between stations~6 and~7 (Figure~\ref{fig:firstRisePeak}\textit{(a,c)}). Both apex-to-conduit walls recover over the same respective intervals, although their pressure increases are smaller. These local increases in front loading support the interpretation of the abrupt conduit-front velocity recoveries as reinforcement of the leading shock when a secondary shock catches it (Figure~\ref{Fig:FrontVel}\textit{(a,b)}). At $H_3$, the circular-wall pressure continues to decrease through the last station at $x=684$~mm. The late velocity recovery in the schlieren record occurs downstream of the last pressure station (Figure~\ref{Fig:FrontVel}\textit{(c)} and Figure~\ref{fig:firstRisePeak}\textit{(e)}).

In the cavity, increasing the height reduces $P_{fp}$ at station~1 for every geometry and shifts the region of greatest front loading downstream. At $H_1$, all six walls reach their largest measured mean peak at station~2 or~3. At $H_3$, all six reach it at station~4 or~5 (Figure~\ref{fig:firstRisePeak}\textit{(b,f)}). The circular wall provides a useful comparison with the optical measurements. At $H_2$, its cavity peak reaches $1.11$~bar at station~3 and falls to $0.56$~bar by station~7. The square wall reaches its largest measured value farther downstream, at station~6 (Figure~\ref{fig:firstRisePeak}\textit{(d)}). This ordering is consistent with the earlier completion of the triple-point crossing for the circular wall and the longer region of Mach-stem growth for the square wall (Figure~\ref{Fig:TriplePoint}\textit{(a)}).

Figure~\ref{fig:firstRiseFraction} compares the first-rise peak pressure with the late-time pressure $P_L$ at cavity station~1. For each shot, $P_L$ is averaged from $2.5$ to $3.5$~ms after local front arrival, and the ratio $P_{fp}/P_L$ is then averaged over shots for each geometry and height. All six walls have their lowest mean ratio at $H_2$, although the lowest first-rise peak pressure occurs at $H_3$ for every wall. The distinction is clearest for the circular wall. At $H_3$, its mean first-rise peak is only $0.55$~bar, yet it exceeds the late-time pressure by approximately $35\%$, the largest relative overshoot among the six geometries at this height.

For the aligned apex-to-cavity wall, the mean first-rise peak pressure at station~1 falls from $0.88$~bar at $H_1$ to $0.51$~bar at $H_2$ and $0.35$~bar at $H_3$. The corresponding mean ratios are $1.02$, $0.92$, and $1.00$. Thus, at $H_3$, even the comparatively weak first rise reaches the late-time pressure level on average. The subsequent drop and gradual recovery seen in Figure~\ref{Tcavity_H}\textit{(f)} explain why the pressure takes considerably longer to settle at that level. This distinction between the magnitude of the first rise and the time required for pressure build-up motivates the comparison with the head-on calibration in the following subsection.

\begin{figure*}[t]
    \centering
    \includegraphics[width=\textwidth]{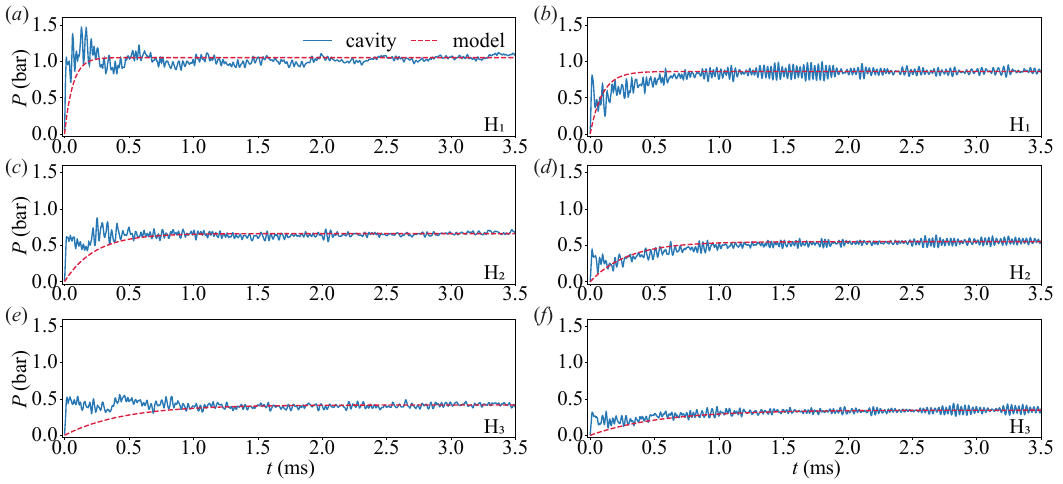}
    \caption{Measured cavity pressure histories at station~1 (solid blue) and model predictions using the head-on calibration (dashed red). Panels \textit{(a,c,e)} show the circular wall and panels \textit{(b,d,f)} show the aligned apex-to-cavity wall. Rows correspond to cavity heights $H_1$, $H_2$, and $H_3$ from top to bottom. Time is measured from cavity-front arrival at station~1. Each prediction uses the head-on coefficient $\alpha$ for the corresponding wall, with the amplitude $F$ set to the measured late-time cavity pressure.}
    \label{fig:Model}
\end{figure*}

\subsubsection{Transferability of the head-on calibration}
\label{sec:transfer}

In the head-on experiments, the arrival of the transmitted shock produced a pressure jump at the end wall, followed by a gradual build-up with superimposed reverberations (Figure~\ref{Headon_P_tau}). The coefficient $\alpha$ obtained for each wall in Section~\ref{sec:Head-onCalibration} characterizes the timescale of this build-up. Under parallel propagation, the pressure at the cavity outer wall also rises abruptly at front arrival, but its subsequent evolution depends on wall geometry and cavity height. We examine whether the head-on calibration describes this evolution at station~1, where the pressure histories show a gradual recovery for the aligned apex-to-cavity wall.

For every geometry and cavity height, the first-rise peak pressure at the cavity outer wall increases from station~1 to station~2 (Figure~\ref{fig:firstRisePeak}). Using station~1 therefore allows us to examine the pressure build-up before this downstream increase in front loading. We apply Eq.~\eqref{eq:Ram_Model} with the cavity height $H$ in place of the head-on stand-off distance. Retaining the head-on coefficient $\alpha$ and the effective length $L_{\mathrm{eff}}$ for each geometry gives
\begin{equation}
    P(t)=F\left[1-\exp\left(-\frac{t}{\tau_{\mathrm{HO}}(H)}\right)\right],
    \qquad
    \tau_{\mathrm{HO}}(H)=\frac{(H+L_{\mathrm{eff}})^\gamma}{\alpha},
    \label{eq:transfer}
\end{equation}
where $t$ is measured from front arrival at the cavity transducer. The amplitude $F$ is the measured late-time cavity pressure $P_L$, averaged from $2.5$ to $3.5$~ms after arrival in each record. The comparison therefore tests the predicted time dependence of the pressure rise. Agreement at late times follows from the prescribed amplitude.

The most direct comparison with the calibration is at $H_2$, where the cavity height equals the $39$~mm stand-off distance of the head-on experiments. For the circular wall, the predicted time is approximately $234~\mu$s. The cavity pressure already approaches its late-time level on the first rise and subsequently fluctuates about that level, while the predicted pressure is still rising (Figure~\ref{fig:Model}\textit{(c)}). For the aligned apex-to-cavity wall, the first peak also approaches the late-time level, reaching on average $92\%$ of $P_L$ (Figure~\ref{fig:firstRiseFraction}). The pressure then drops and recovers over several hundred microseconds. The head-on prediction follows the timescale of this recovery more closely, although the exponential response cannot reproduce the initial peak and drop (Figure~\ref{fig:Model}\textit{(d)}).

The distinction between the two walls is particularly clear at $H_3$. The circular wall reaches a mean first-rise peak of $1.35P_L$, and the aligned apex-to-cavity wall reaches approximately $P_L$ (Figure~\ref{fig:firstRiseFraction}). For the circular wall, the model remains below the measured cavity pressure over much of the first millisecond (Figure~\ref{fig:Model}\textit{(e)}). For the aligned apex-to-cavity wall, the drop following the first peak again leaves a gradual recovery toward $P_L$ (Figure~\ref{fig:Model}\textit{(f)}). Thus, the magnitude of the first peak alone does not explain the difference in model agreement. The important distinction is how the pressure evolves after that peak.

To quantify the comparison for all six geometries, we fit Eq.~\eqref{eq:transfer} over the first $1.5$~ms after cavity-front arrival, retaining the measured $F$ and allowing $\alpha$ to vary. The resulting $\tau_{\mathrm{free}}=(H+L_{\mathrm{eff}})^\gamma/\alpha_{\mathrm{free}}$ is an effective time for the measured pressure history over this interval. Table~\ref{tab:T_ratio} gives the median ratio $\tau_{\mathrm{free}}/\tau_{\mathrm{HO}}$ over repeated experiments. Values above unity indicate a slower fitted response than predicted, and values below unity indicate a faster response.

\begin{table}[!ht]
    \centering
    \caption{Ratio of the fitted cavity pressure rise time $\tau_{\mathrm{free}}$ to the rise time $\tau_{\mathrm{HO}}$ predicted from the head-on calibration at station~1. Values are medians over repeated experiments. A dagger indicates that the fitted rise time reached its lower bound.}
    \label{tab:T_ratio}
    \setlength{\tabcolsep}{3.5pt}
    \begin{tabular}{@{}lccc@{}}
        \hline
        & \multicolumn{3}{c}{$\tau_{\mathrm{free}}/\tau_{\mathrm{HO}}$} \\
        Internal geometry & $H_1$ & $H_2$ & $H_3$ \\
        \hline
        Circular                 & 0.096 & 0.043 & $0.025^{\dagger}$ \\
        Square                   & 0.22  & 0.29  & 0.044              \\
        Triangular, apex-to-conduit & 0.16  & 0.13  & $0.038^{\dagger}$ \\
        Triangular, apex-to-cavity  & 1.9   & 1.3   & 0.61               \\
        Triangular, apex-to-conduit, staggered & 0.24  & 0.15  & $0.028^{\dagger}$ \\
        Triangular, apex-to-cavity, staggered & 0.17  & 0.43  & $0.021^{\dagger}$ \\
        \hline
    \end{tabular}
\end{table}

The two apex-to-cavity arrangements show how the effect of geometry changes between loading configurations. Under head-on impingement, these walls have the two longest characteristic times, approximately $290~\mu$s for the aligned wall and $322~\mu$s for the staggered wall (Table~\ref{tab:headon}). At the matching height $H_2$ under parallel loading, their fitted times are approximately $368$ and $138~\mu$s, respectively. Their ordering therefore reverses. This difference is consistent with the more pronounced pressure drop and subsequent recovery observed for the aligned wall in the station-1 histories (Figure~\ref{Parallel_H2}\textit{(h,l)}).

For the aligned apex-to-cavity wall, $\tau_{\mathrm{free}}/\tau_{\mathrm{HO}}$ decreases from $1.9$ at $H_1$ to $1.3$ at $H_2$ and $0.61$ at $H_3$. Agreement is closest at $H_2$, the original calibration distance. From $H_2$ to $H_3$, the head-on calibration predicts an increase in rise time from approximately $290$ to $537~\mu$s, whereas the fitted time decreases from approximately $368$ to $327~\mu$s.

For the other five geometries, the fitted response is faster than the head-on prediction at every height, with $\tau_{\mathrm{free}}/\tau_{\mathrm{HO}}$ ranging from $0.021$ to $0.43$. At $H_3$, four of these geometries reach the lower fitting bound on $\tau_{\mathrm{free}}$, which limits the resolution of the shortest times. For the aligned apex-to-cavity wall, the gradual recovery after the first pressure peak is evident in the station~1 histories, and the effective rise time fitted over the first $1.5$~ms remains comparable to the head-on prediction at all three heights. For the other five walls, the head-on calibration systematically overpredicts the effective rise time under parallel loading.

\raggedbottom
\section{Conclusions}
\label{sec:Conclusions}
We investigated how the internal geometry of a porous wall and the height of an adjacent cavity affect shock propagation and pressure loading as a shock travels along the wall. Schlieren recordings and pressure measurements for six wall geometries showed that increasing the cavity height delayed front arrival at the cavity outer wall and extended the downstream region over which the cavity front developed. At a given height, the wall geometry influenced how far the conduit shock traveled before the cavity front became nearly normal and whether the cavity front subsequently overtook it.

Increasing the cavity height from $9$ to $69$~mm increased the cavity front's arrival lag near the entrance from approximately $60$ to $185$~$\mu\mathrm{s}$. In the taller cavity, the transmitted disturbances traveled farther before reaching the outer wall, and the Mach stem grew across the cavity over a longer downstream distance. The pressure measurements showed a corresponding shift in the location of peak loading. For all six geometries, the largest measured cavity pressure peak at front arrival occurred $159$--$264$~mm downstream of the entrance in the shallowest cavity and $369$--$474$~mm downstream in the tallest.

The $39$~mm cavity showed how the evolution of both fronts determined their arrival order. For the circular elements, the Mach stem spanned the cavity closest to the entrance, and the cavity front was the first among the six geometries to overtake the conduit shock. Farther downstream, the conduit shock regained speed and retook the lead. Its pressure peak rose from $0.32$ to $0.70$~bar between the final two measurement locations, consistent with reinforcement by a secondary shock. For the square elements and the aligned triangles with their apices facing the cavity, the Mach stem spanned the cavity farthest downstream. Despite this similarity, the cavity front overtook the conduit shock near $x=815$~mm for the square elements, whereas it remained behind the conduit shock throughout the observed region for the aligned triangles.

The head-on experiments isolated the effect of passage orientation at fixed void fraction and frontal blockage. Reversing the aligned triangles so that the flow encountered the narrow opening first increased the characteristic pressure-rise time from approximately $175$ to $290$~$\mu\mathrm{s}$. Simulations showed flow separation at the entrance to the widening passages and later emergence of the transmitted front. This delayed transmission is consistent with the longer distance required for the cavity front to become nearly normal under parallel loading. The effect of staggering depended on the loading direction. With the triangular apices facing the cavity, staggering increased the head-on rise time from $290$ to $322$~$\mu\mathrm{s}$. At the matching cavity height of $39$~mm, it reduced the effective rise time near the entrance under parallel loading from $368$ to $138$~$\mu\mathrm{s}$. Thus, changing the direction of shock propagation relative to the wall reversed the ordering of the two geometries by pressure-rise time.

From an engineering perspective, the wall geometry and cavity height can be chosen to shape the pressure-loading profile inside the cavity during the early stages of shock propagation. They influence how long the initial load is delayed and how far downstream the transmitted front develops. The benefit of a weaker initial pressure rise is therefore limited in both time and distance. Once the Mach stem spans the cavity, the nearly normal shock produces a sharp pressure jump at the outer wall. At downstream locations, this jump can equal or exceed the conduit pressure jump at the same position. The pressure history at a component inside the cavity therefore depends on where it lies relative to the region in which the transmitted shock develops.

\section*{CRediT authorship contribution statement}
\textbf{Omri Ben Haim:} Methodology, Investigation, Formal analysis, Data curation, Software, Visualization, Writing -- original draft. \textbf{Yoav Gichon:} Methodology, Investigation. \textbf{Raz Heppner:} Software, Investigation, Visualization, Writing -- review \& editing. \textbf{Omri Ram:} Conceptualization, Methodology, Supervision, Resources, Funding acquisition, Writing -- review \& editing.

\section*{Declaration of competing interest}
The authors declare that they have no known competing financial interests or personal relationships that could have appeared to influence the work reported in this paper.

\section*{Data availability}
Data will be made available on request.

\section*{Declaration of generative AI and AI-assisted technologies in the writing process}
During the preparation of this manuscript, the authors used OpenAI ChatGPT (GPT-5.5 and 5.6) and Anthropic Claude (Claude Opus 5.5) to assist with language editing, to improve the clarity and structure of the text, and to help format the LaTeX source. The authors reviewed and edited all AI-assisted output and take full responsibility for the content of the publication.

\section*{Acknowledgments}
The results presented here are based upon work supported by the Israel Science Foundation grant 3113/21 and by the Zuckerman STEM Leadership Program. We thank Dr. Jibu Tom Jose for proofreading the manuscript, assisting with the analysis, and offering advice. We also thank Dr. Yigal Evron for his technical assistance in conducting the experiments and for his advice and support throughout the study.

\bibliographystyle{elsarticle-num}
\bibliography{cas-refs_review_2026-11-20}

\end{document}